\documentclass[journal]{IEEEtran}

\usepackage{cite}

\ifCLASSINFOpdf
   \usepackage[pdftex]{graphicx}
\else
\fi
\usepackage{amsmath}
\usepackage{dsfont}

\usepackage[utf8]{inputenc}
\usepackage[T1]{fontenc}
\usepackage{lmodern}
\usepackage{subcaption}
\usepackage{bm}
\usepackage{xcolor}

\begin{document}

\title{Data Transmission based on Exact Inverse Periodic Nonlinear Fourier Transform, Part II: Waveform Design and Experiment}

\author{Jan-Willem~Goossens,~\IEEEmembership{Student Member, IEEE,}
        Hartmut~Hafermann,~\IEEEmembership{Senior Member, IEEE}
        and~Yves~Jaou\"en~\IEEEmembership{}%
\thanks{J.-W.~Goossens and H.~Hafermann are with the Optical Communication Technology Lab, Paris Research Center, Huawei Technologies France, 92100 Boulogne-Billancourt, France}%
\thanks{J.-W.~Goossens and Y.~Jaou\"en are with LTCI, T\'el\'ecom Paris, Universit\'e Paris-Saclay, 91120 Palaiseau, France}%
}

\maketitle

\begin{abstract}

The nonlinear Fourier transform has the potential to overcome limits on performance and achievable data rates which arise in modern optical fiber communication systems when nonlinear interference is treated as noise.
The periodic nonlinear Fourier transform (PNFT) has been much less investigated compared to its  counterpart based on vanishing boundary conditions.
In this paper, we design a first experiment based on the PNFT in which information is encoded in the invariant nonlinear main spectrum. To this end, we propose a method to construct a set of periodic waveforms each having the same fixed period, by employing the exact inverse PNFT algorithm developed in Part I.
We demonstrate feasibility of the transmission scheme in experiment in good agreement with simulations and obtain a bit-error ratio of $10^{-3}$ over a distance of 2000 km. It is shown that the transmission reach is significantly longer than expected from a naive estimate based on group velocity dispersion and cyclic prefix length, which is explained through a dominating solitonic component in the transmitted waveform.
Our constellation design can be generalized to an arbitrary number of nonlinear degrees of freedom.
\end{abstract}

\IEEEpeerreviewmaketitle

\section{Introduction}

\IEEEPARstart{F}{ueled} by ever increasing online traffic demands, the capacity of optical communication links has grown exponentially in a Moore-like fashion, following a tenfold increase every four years~\cite{Richardson2013}. This trend has been kept up owing to the  introduction of new technologies, starting from improved transmission fibers and continuing with the Erbium-doped fiber amplifier, wavelength-division multiplexing and high spectral efficiency coding. Even with advanced digital signal processing and coded modulation schemes, in particular probabilistic shaping~\cite{Bocherer2019}, modern optical transmission systems are approaching the so-called nonlinear Shannon limit~\cite{Ellis2010}.
Not regarded as fundamental~\cite{Secondini2016}, it emerges as an artifact of the application of communication techniques developed for linear channels, in which nonlinear interference is treated as noise. While SNR increases limitless with increasing power, the signal-to-interference-plus-noise ratio reaches an optimum~\cite{Essiambre2010}.

Recently, optical communication techniques based on the nonlinear Fourier transform have reentered the focus in the search for a candidate technology to overcome these limitations (for a review, see Ref.~\cite{Turitsyn2017}).
After the introduction of eigenvalue communication~\cite{Hasegawa1993}, the idea of encoding information in the nonlinear spectrum has been extended from the discrete to the continuous nonlinear spectrum in nonlinear frequency division multiplexing (NFDM)~\cite{Yousefi2014} and the nonlinear inverse synthesis (NIS) method~\cite{Le2014}.
These approaches hinge on the property of the nonlinear modes to propagate independently, which at least in principle overcomes the problem of nonlinear interference in optical networks.

The conventional NFT assumes vanishing boundary conditions, which requires systems to operate in burst mode. Signal bursts are separated by guard intervals long enough to avoid overlap caused by signal broadening due to fiber dispersion.
To increase the spectral efficiency it is desirable to decrease the size of the guard interval, or to increase the burst length.
It has been shown that pre-compensation halves the necessary guard interval and that processing complexity at the receiver can be reduced using a windowing technique. 
However there are indications that signal-noise interactions accumulating in the forward NFT for vanishing boundary conditions  degrade performance, so that spectral efficiency cannot be increased at will by increasing burst length~\cite{Civelli2017}.
At the same time, the development of fast and accurate algorithms for the forward nonlinear Fourier transform continues~\cite{Wahls2015,Vaibhav2018,Chimmalgi2018}.

A number of works have studied and improved practical aspects of the approach. This includes the application to fiber links with periodic lumped amplification~\cite{Kamalian2017} or the generalization to polarization multiplexing to double the spectral efficiency~\cite{Goossens2017,Gaiarin2018,Civelli2018}. 
It has been shown that with comparable resources of time and bandwidth, NFDM can outperform OFDM in terms of achievable information rates~\cite{Yousefi2020}.
The improvements have been implemented in a number of experiments~\cite{Le2017,Le2017b,Le2018,Aref2018, Gaiarin2018a,Gui2018,Gui2018a}. To date, a record experimental data rate of 220 Gb/s for a polarization-multiplexed system based on b-modulation has been reported, achieving a spectral efficiency of 4 bits/s/Hz~\cite{Yangzhang2019}.

The periodic NFT has been less the subject of study.
This is largely due to the fact that the mathematics underlying the periodic NFT is significantly more complex compared to the conventional NFT.
The underlying mathematical structure however might provide efficient means to compute the nonlinear superposition of signals required for multiplexing, for example through algebro-geometric reduction~\cite{Belokolos1994,Smirnov2013}.
Another advantage of this algebro-geometric approach to the inverse PNFT utilized in this paper is that it provides exact solutions in the form of analytical expressions. This can be exploited in the signal design as detailed below. Periodic signals also have a well-defined time duration. While they do require a cyclic prefix (CP) to avoid inter-symbol interference (ISI), the processing window is independent of the channel memory and limited to a single period~\cite{Turitsyn2017}.
Stable and fast numerical algorithms are also known for the periodic forward transform~\cite{Wahls2015,Kamalian2016}.

As already detailed in Part I~\cite{GoossensI}, the inverse PNFT for a given nonlinear spectrum leads to exact solutions which however are not in closed form: Their parameters are given in terms of algebro-geometric loop integrals over a Riemann surface of a certain genus~\cite{Kotlyarov2014,Tracy1984,Tracy1988,Belokolos1994}.
For certain nearly degenerate cases, periodic solutions with a defined nonlinear spectrum are known exactly and in closed form. 
Data transmission without the need to evaluate Riemann theta functions or to compute loop integrals can be achieved using perturbed plane waves~\cite{Kamalian2016a}. Transmission of solutions computed in terms of simpler one-dimensional elliptic functions using algebraic reduction~\cite{Belokolos1994} has been demonstrated in numerical simulations~\cite{Kamalian2018c}.

The perturbative character of the signals however limits the size of the constellation and requires equalization at the receiver.

To avoid such restrictions, the inverse transform must be computed explicitly. For the special case of genus 2, the calculation can be simplified through algebro-geometric reduction but the generalization to higher genus is not straightforward.
A complementary approach is to compute the inverse PNFT through the solution of a Riemann-Hilbert problem~\cite{Kamalian2018}. It has the advantage that it reduces the complexity of the inverse transform step compared to the algebro-geometric approach~\cite{Kamalian2018b}. To obtain periodic solutions, an optimization problem has to be solved~\cite{Kopae2020}. In the algebro-geometric approach, one can exploit the availability of an analytic expression.

In this paper, we design an experiment based on the periodic NFT. While the computation of loop integrals may be too complex for online computation, we will pre-compute the information carrying signals and store them in lookup-tables at the transmitter. Feasibility of transmission based on this scheme has been demonstrated in Ref.~\cite{Goossens2019}. Here we extend this work by exploring the factors that influence constellation design in a PNFT based transmission system.
We propose a general method based on the exact inverse PNFT algorithm developed in Part I~\cite{GoossensI} to generate a set of periodic solutions with identical period, which are approximately finite-gap and suitable for information transmission. It is shown that the solitonic character of the signals significantly increases transmission reach compared to the expectation based on the assumption of a linear dispersive channel given the CP length.

The organization of the paper is as follows: In Sec.~\ref{sec:pnft} we briefly introduce the mathematical formalism underlying the PNFT. Our signal design is based on the algebro-geometric integration method to compute the exact inverse PNFT, which we outline in Sec.~\ref{sec:ipnft}. The section summarizes the main results of Part I to make the paper self-contained.
In this section we also discuss approaches to obtain quasi-periodic solutions. Sec.~\ref{sec:communication} provides the details of our PNFT transmission scheme, in particular our proposed method to obtain constellations of signals with the same period and the construction of the transmitted waveform. 
We demonstrate the feasibility of the proposed communication scheme in experiment. Sec.~\ref{sec:experiment} describes the experimental setup and Sec.~\ref{sec:results} the experimental and numerical results.
Sec.~\ref{sec:discussion} concludes the paper.

\section{Periodic nonlinear Fourier transform}
\label{sec:pnft}

The periodic nonlinear Fourier transform hinges on the applicability of an integrable model for light propagation in optical fiber.
The integrable, self-focusing nonlinear Schr\"odinger equation describing the spacetime evolution of the complex envelope $q(t,z)$ of an optical carrier is given in its dimensionless form by~\cite{Agrawal2000}
\begin{equation}
i\frac{\partial q(t,z)}{\partial z} + \frac{\partial^2 q(t,z)}{\partial t^2} + 2|q(t,z)|^2 q(t,z) = 0.
\label{eq:nlse}
\end{equation}
Here $z$ is the space coordinate which measures the distance along the fiber and $t$ is the retarded time.

There are two main factors that break integrability of a fiber link in a real communication system: fiber losses and noise. While losses due to fiber attenuation can at least in principle be completely compensated by ideal distributed Raman amplification~\cite{Agrawal2000}, it requires precise control of multiple pumps and is difficult to achieve in practice~\cite{Ania2004}.
Here we consider practical multi-span fiber links with repeated lumped amplification based on Erbium-doped fiber amplifiers (EDFAs). In this case the NLSE still serves as an approximate model as long as the signal envelope varies slowly over the distance of a fiber span~\cite{Le2015}.
Noise in this scenario mainly stems from the amplifiers in the form of amplified spontaneous emission (ASE) noise. As noise is inevitable, we will base our communication scheme on the common assumption that it can be regarded as a small perturbation of the integrable NLSE.

As for the case of vanishing boundary conditions, the formulation of the NFT for periodic boundary conditions starts from the Zakharov-Shabat (ZS) system~\cite{Shabat1972}, which can be stated as
\begin{equation}
\frac{\partial}{\partial t}\Phi(t,z,\lambda) = \left[-i\lambda\sigma_3 + \left(\begin{array}{cc}0 & -q(t,z)\\ \bar{q}(t,z) &0\end{array}\right)\right]\Phi(t,z,\lambda).
\label{eq:zs}
\end{equation}
Obtaining the scattering data $\Phi(t,z,\lambda)$ from the waveform $q(t,z)$ is known as the ZS scattering problem. The waveform $q(t,z)$ hereby plays the role of a scattering potential.
We are interested in periodic potentials. In this case the two-component vectors $\Phi(t,z,\lambda)$ are called Bloch functions.
The main spectrum of the PNFT for the NLSE is given by those eigenvalues $\lambda$ in the Zakharov-Shabat system for which the Bloch functions are (anti-)periodic in time.
Since~\eqref{eq:zs} can be recast into the form of an eigenvalue problem $L \Phi=\lambda \Phi$, the values $\lambda_j$ are referred to as eigenvalues. They come in complex conjugate pairs. We therefore specify the main spectrum solely in terms of its eigenvalues with positive imaginary part.
The main spectrum has the important property that it remains invariant while in general the waveform evolves in a complicated way. This property obviously makes the main spectrum interesting for encoding information. 

The complete description of the PNFT spectrum requires additional auxiliary spectrum variables $\mu_j(t,z)$ which govern the spacetime evolution of the waveform. Their definition is not unique~\cite{Ma1981, Kotlyarov2014} and they evolve according to a set of complicated coupled nonlinear partial differential equations~\cite{Kotlyarov2014}. Here we restrict ourselves to encoding information in the invariant main spectrum.

\section{Exact inverse PNFT}
\label{sec:ipnft}

We design constellations of finite-gap solutions constructed by means of the algebro-geometric integration method first introduced in~\cite{Kotlyarov2014}. A thorough review of the algebro-geometric approach to the solution of nonlinear integrable equations is provided in \cite{Belokolos1994}. We compute our solutions with the algorithm from Part I~\cite{GoossensI}.
Finite-gap solutions are characterized by a main spectrum that consists of a finite number of points~\cite{Kotlyarov2014}.
They can be expressed in terms of a ratio of Riemann theta functions in the form
\begin{equation}
q(t,z) = K_0\frac{\theta\left(\frac{1}{2\pi}(\underline{\omega}t+\underline{k}z+\underline{\delta}^-)|\bm{\tau}\right)}{\theta\left(\frac{1}{2\pi}(\underline{\omega}t+\underline{k}z+\underline{\delta}^+)|\bm{\tau}\right)} e^{i\omega_0 t+ik_0z},\label{eq:thetasol}
\end{equation}
where the $g$-dimensional Riemann theta function is given by
\begin{equation}
\theta(\underline{x}|\bm{\tau}) = \sum_{m_1=-\infty}^\infty \ldots\sum_{m_g=-\infty}^\infty \exp(\pi i\underline{m}^T\bm{\tau}\underline{m}+ 2\pi i\underline{m}^T\underline{x}).\label{eq:theta}
\end{equation}
Here $K_0$, $k_0$ and $\omega_0$ are scalars, $\underline{k}, \underline{\omega}, \underline{\delta}^-$ and $\underline{\delta}^+$ are $g$-dimensional vectors and $\bm{\tau}$ is the $g\times g$ period matrix.
In the algebro-geometric integration method, these parameters are computed as loop integrals over a two-sheeted Riemann-surface of genus $g$ of an algebraic curve which is defined in terms of the $2g+2$ eigenvalues of the main spectrum (see Part I for details),
$$
\Gamma : \left\{ (P,\lambda), P^2 = \prod_{k=1}^{g+1} (\lambda-\lambda_k)(\lambda-\bar{\lambda}_k),P,\lambda\in\mathds{C}\right\}.
$$
As an example, the frequency vector $\underline{\omega}$ is given by 
\begin{equation}
\omega_j = -4\pi i (\bm{A}^{-1})_{j,g}\textrm{, with }(\bm{A})_{j,k} = \int_{a_k}dU_j,
\label{eq:periods}
\end{equation}
where the $dU_j = (\lambda^{j-1}/P(\lambda)) d\lambda$ for $j=1,\ldots g$ are a basis of holomorphic differentials on $\Gamma$ and the $a_k$ are closed paths on the Riemann surface and defined as part of a canonical homology basis of cycles~\cite{Belokolos1994,GoossensI}.

The only parameters in \eqref{eq:thetasol} that depend on the auxiliary spectrum $\mu_j(t,z)$ are the  phases $\underline{\delta}^\pm$. However, their difference 
\begin{equation}\underline{\delta}^+-\underline{\delta}^- = \bm{A}^{-1}\int_{\infty^-}^{\infty^+} d\underline{U}\end{equation} is independent of the auxiliary spectrum. Furthermore, it can be proven that any vector $\underline{\delta}^+$ corresponding to a solution of the NLSE is real, and all solutions of the NLSE  correspond to a purely real vector $\underline{\delta}^+$~\cite{Belokolos1994}, see also the Appendix of Part I.
Since in this paper we do not encode information in the auxiliary spectrum, we implicitly choose the auxiliary spectrum by fixing $\underline{\delta}^+=\underline{0}$. 

\subsection*{Quasi-periodic solutions}

The period of a finite gap solution is determined by the frequencies $\omega_j$, $j=1,\ldots,g$ and $\omega_0$. The Riemann theta function, Eq. \eqref{eq:theta}, is periodic in each component of the vector argument $\underline{x}$ with period 1. Therefore the ratio of theta functions is periodic in time with period $T_p<\infty$ when $\omega_j T_p/(2\pi)$ is an integer for $1\leq j\leq g$. Such a $T_p$ only exists when all pairs $\omega_i, \omega_j$ are commensurate, i.e. their ratio is a rational number; otherwise the time-period is infinite. 
When in addition $\omega_0$ is commensurate with all $\omega_j$, the solution~\eqref{eq:thetasol} is periodic. Otherwise it is quasi-periodic, that is, periodic up to a phase $\omega_0 T_p$. In this case we call $T_p$ the quasi-period.

For our purposes it is sufficient to construct solutions that are quasi-periodic. 
The main spectrum of a quasi-periodic solution can be recovered through a forward PNFT (FPNFT) algorithm designed for periodic signals~\cite{Wahls2017,Wahls2018} by means of the identity
\begin{equation}
\textrm{FPNFT}[q(t)e^{i\phi t/T_p}] = \textrm{FPNFT}[q(t)]-\frac{\phi}{2T_p},\label{eq:quasi}
\end{equation}
which relates a constant shift $\Lambda=-\phi/(2T_p)$ of the main and auxiliary spectra, $\lambda_k\to\lambda_k+\Lambda$ and $\mu_j(t,z)\to\mu_j(t,z)+\Lambda$ to a change in phase $\exp(-2i\Lambda t)$ of the solution $q(t)$ (see Appendix B of Part I~\cite{GoossensI}).
To this end, the phase $\phi$ is chosen according to $\omega_0T_p = \phi \text{ mod } 2\pi$ to make $q(t)e^{i\phi t/T_p}$ periodic. The forward PNFT is computed of the periodic waveform and the inverse frequency shift is applied in the nonlinear Fourier domain.
Since $\phi$ is unknown at the receiver it has to be estimated based on the phase difference between start and end of the waveform.
Apart from this technicality, the difference between periodic and quasi-periodic solutions of the NLSE is largely artificial: When a solution is modulated onto the carrier, a periodic solution of the NLSE in general corresponds to a quasi-periodic physical field and possibly vice-versa.

In general, obtaining quasi-periodic solutions requires inverting the relation~\eqref{eq:periods} to obtain the main spectrum. Finding a main spectrum such that the frequencies $\omega_j$ satisfy a commensurability constraint is a difficult problem~\cite{Belokolos1994}.
For the particular case of genus 2, it is possible to guarantee exactly time-periodic solutions by imposing constraints on the \emph{symmetry} of the nonlinear spectrum.
For a main spectrum of the form $\{-a+bi, ci, a+bi\}$ or $\{ai, bi, ci\}$ with $a,b,c\in \mathds{R}$ the solution is periodic in time~\cite{Smirnov2013, Smirnov2014}: there exist two holomorphic involutions (negation and complex conjugation) on the genus-2 surface, which implies that it covers two genus-1 surfaces. The elements of the homology basis of cycles on the genus-2 surface can therefore be represented as linear combinations of cycles on two genus-1 surfaces with integer coefficients. It follows that the frequencies and wave vectors in~\eqref{eq:thetasol} are integer multiples of each other. For spectra of the form $\{-a+bi, ci, a+bi\}$ this yields $k_1 = 2k_2$, $\omega_1 = 0$ and $\omega_2 = \omega$. Since in addition $\omega_0=0$, the solution is strictly periodic.
The reduction of Abelian integrals and Riemann theta functions to lower genera generalizes to surfaces of higher genus. It is however not obvious how to exploit it to construct periodic solutions of higher genus.

\section{PNFT based signal design}
\label{sec:communication}

Our goal is to design a communication system based on the PNFT where the data-carrying signal will consist of a train of symbols, each of which consists of a period of a finite-gap solution and a CP to preserve the nonlinear spectrum.
As a proof of principle we design a set of four waveforms having the same fixed finite period. 
We call this set the constellation and each element a symbol. Two bits of information are encoded in each symbol.
Constraining the symbols to the same period greatly simplifies modulation and symbol separation and guarantees a well defined nonlinear spectrum in the receiver digital signal processing (DSP). 
We focus here on genus-2 solutions. The main spectrum of each symbol is therefore fully described by three complex numbers in the upper half plane.
The impact of the degree of freedom used to make all symbols occupy the same period on the data rate is negligible for transmission based on solutions of higher genus.

\begin{figure}
     \centering       
      \begin{subfigure}[t]{0.5\textwidth}
         \raisebox{-\height}{\includegraphics[width=0.9\textwidth]{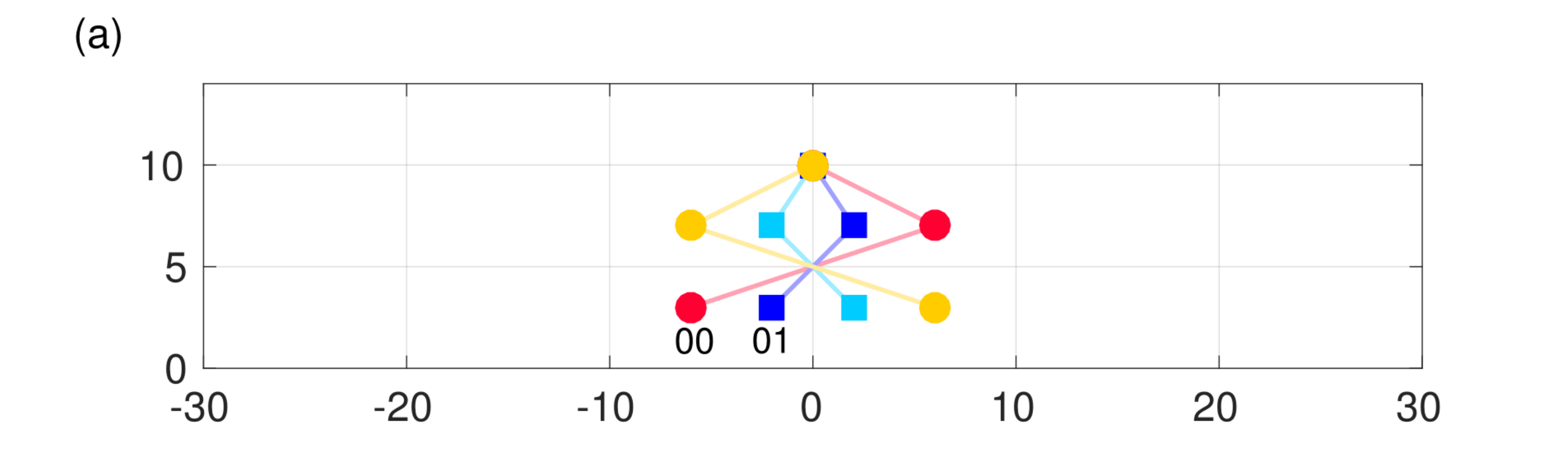}}%
				 \vspace{-0.2cm}
         \raisebox{-\height}{\includegraphics[width=0.9\textwidth]{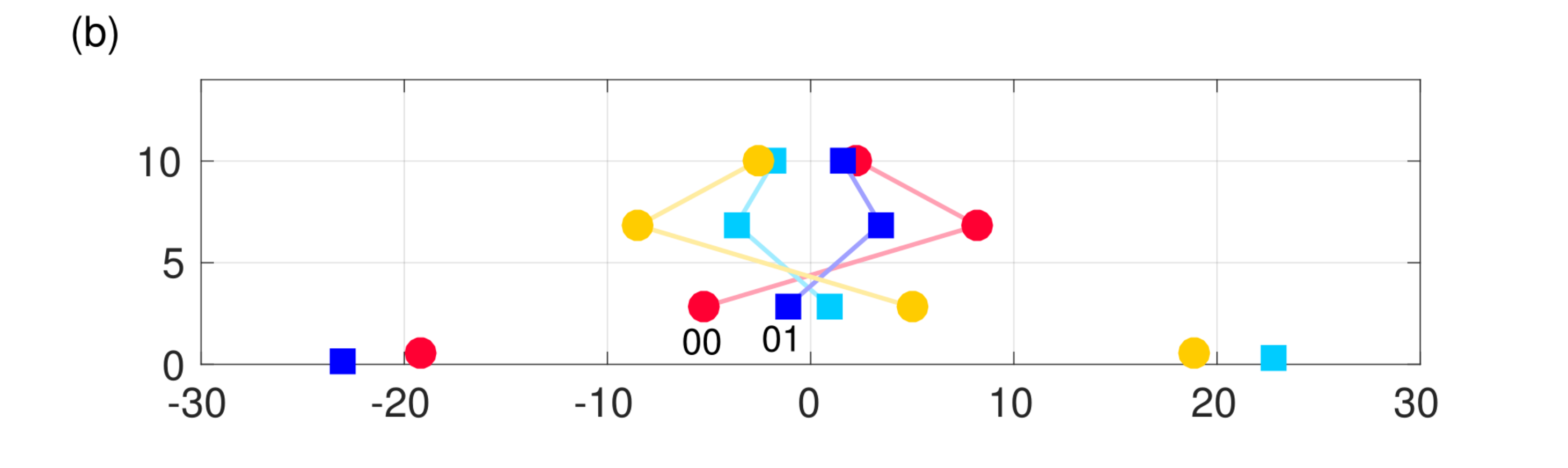}}
				 \raisebox{-\height}{\includegraphics[width=0.9\textwidth]{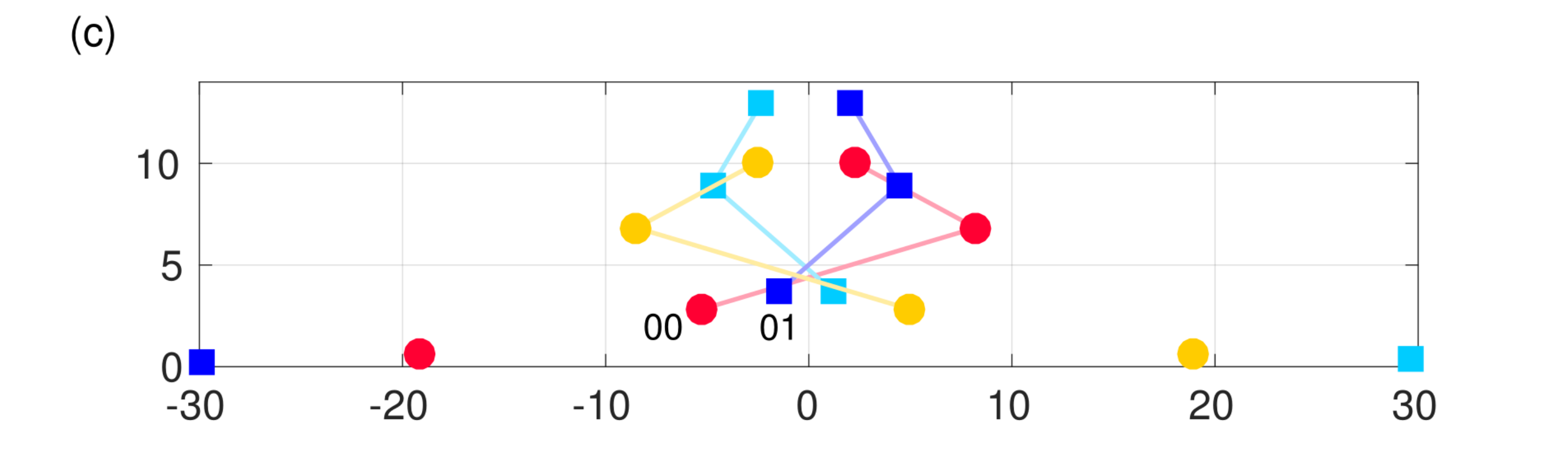}}
         \raisebox{-\height}{\includegraphics[width=0.9\textwidth]{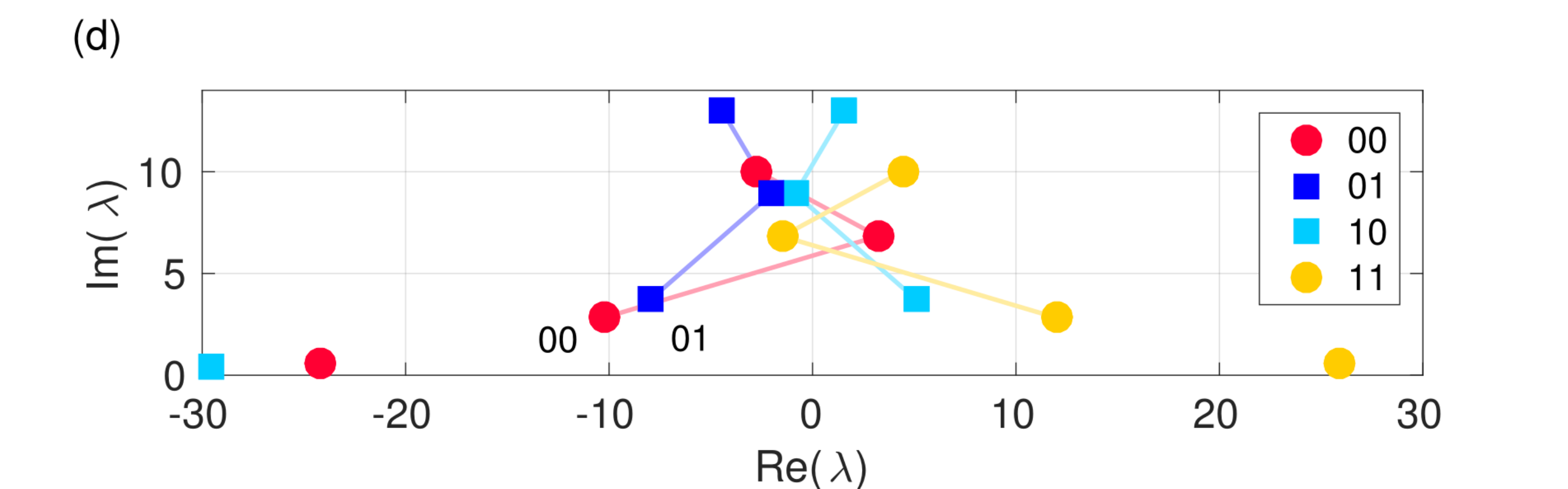}}
    \end{subfigure}
	\caption{(a) Unconstrained nonlinear spectra (b) after setting one of the $\omega_j$ to zero, (c) after making all periods the same, and (d) after fixing the group velocities to the same value.\label{fig:Symbols} }
	\end{figure}

\begin{figure*}
    \begin{subfigure}[t]{0.24\textwidth}
        \raisebox{-\height}{\includegraphics[width=\textwidth]{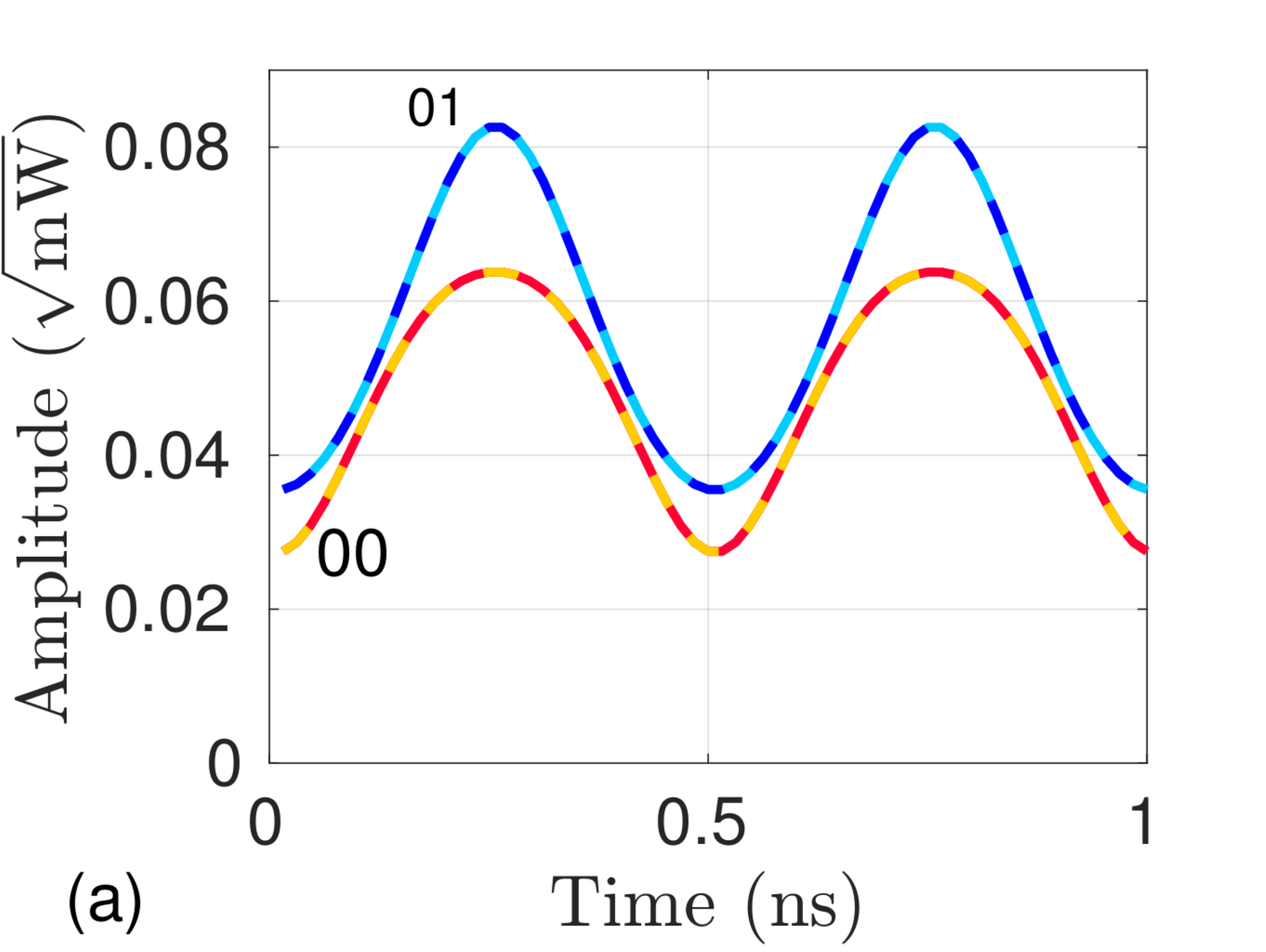}}
		\label{fig:Amplitude}
    \end{subfigure}         
    \begin{subfigure}[t]{0.24\textwidth}
        \raisebox{-\height}{\includegraphics[width=\textwidth]{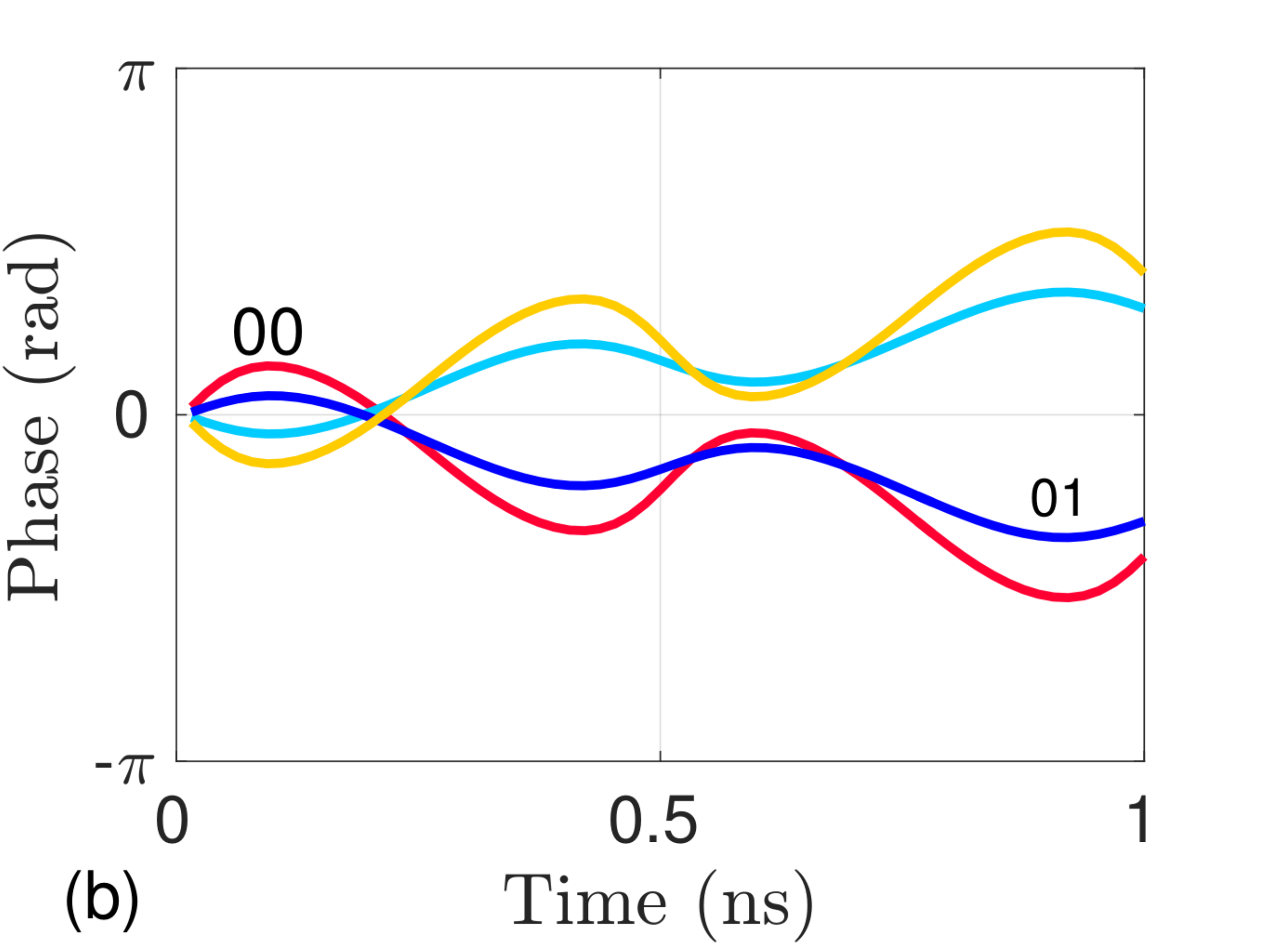}}
		\label{fig:Phase}
    \end{subfigure}
		  \begin{subfigure}[t]{0.24\textwidth}
        \raisebox{-\height}{\includegraphics[width=\textwidth]{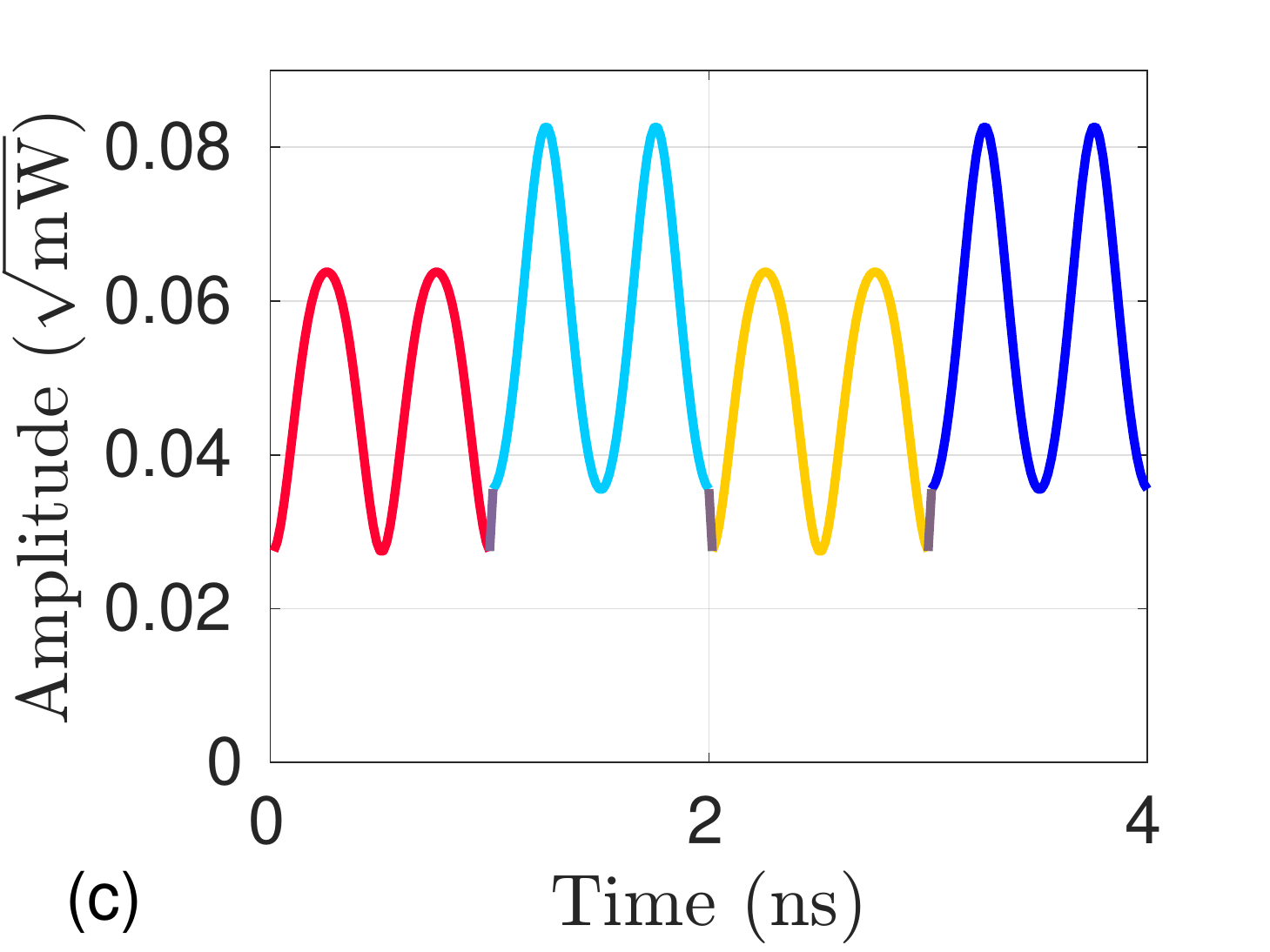}}
		\label{fig:AmplitudeCombined}
    \end{subfigure}
		  \begin{subfigure}[t]{0.24\textwidth}
        \raisebox{-\height}{\includegraphics[width=\textwidth]{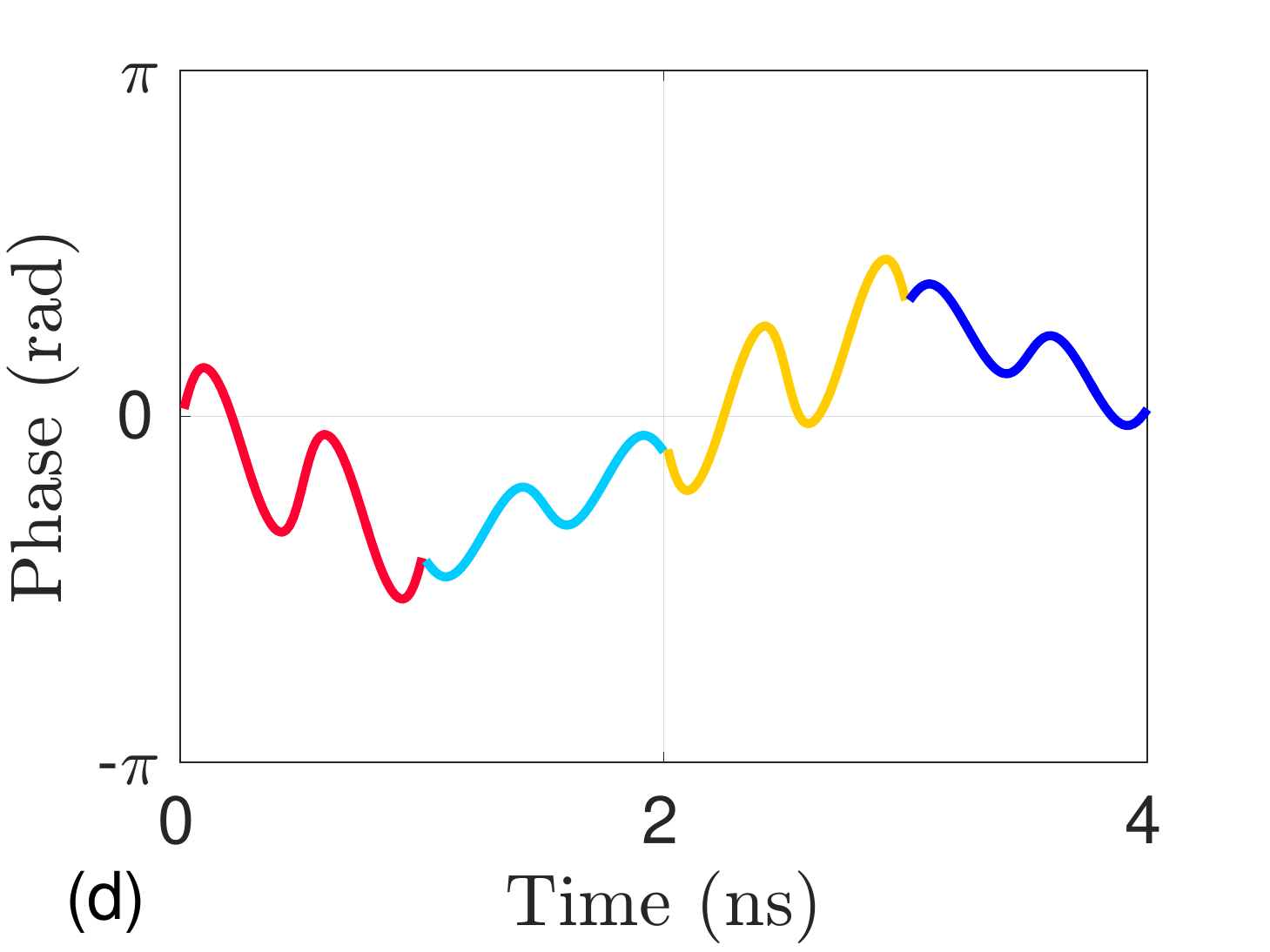}}
		\label{fig:PhaseCombined}
    \end{subfigure}
  \caption{\label{fig:waveforms} Panels (a) and (b) show the amplitude and phase of the symbols of the constellation corresponding to the spectra of Fig.~\ref{fig:Symbols} including the 50\% CP. Panels (c) and (d) illustrate how these are combined to minimize the discontinuity on the boundaries.}
\end{figure*}

\subsection{Method to obtain quasi-periodic solutions}

We obtain the constellation in three steps. First, we obtain a set of quasi-periodic solutions. Second, we match the solutions to have the same period $T_p$. Third, we match the group velocities of the symbols to reduce the size of the required CP.

Our method to construct periodic solutions is motivated by the perturbed plane wave (PPW) solutions of Ref.~\cite{Kamalian2016}.
To construct PPW solutions, one starts from a plane wave which is a periodic genus-0 solution whose main spectrum consists of a single non-degenerate point $ai$ and a degenerate point on the imaginary axis, $bi$. The degenerate point is split in two: $bi\to bi\pm \frac{1}{2}\epsilon$, where $\epsilon\in\mathds{C}$.
The resulting spectrum is genus-2. If the perturbation $\epsilon$ is small, the solution remains approximately periodic\footnote{Note that the terminology is somewhat misleading: although the initial condition for these solutions is a small perturbation of a plane wave, these waveforms do in general \emph{not} remain close to a plane wave during propagation, although the nonlinear spectrum will of course be preserved.}~\cite{Kamalian2016a} (to first order in $\epsilon$), see~\cite{Tracy1988} for a derivation.

In the following we proceed similarly. However we emphasize that the resulting waveforms are no longer simple perturbations of a plane wave. In particular, they can no longer be described by the closed-form expressions of Ref.~\cite{Kamalian2016}, but instead must be computed exactly via the finite-gap integration method of Part I, as outlined in Sec.~\ref{sec:pnft}.
In addition, because exact quasi-periodicity is essential for the recovery of the spectrum, we shift our focus: Instead of constructing a solution which is exactly genus-2 and approximately quasi-periodic, we construct an exactly quasi-periodic solution whose spectrum remains approximately genus-2 in the sense described below.

We start with genus-2 spectra of the form $\{ai, bi+\gamma, bi-\gamma\}$ where $a,b$ are real numbers and $\gamma\in \mathds{C}$ is a complex number, see Fig.~\ref{fig:Symbols} (a).
The value $2|\gamma|$ plays the role of $\epsilon$ in the PPW approach, but here it is not small. 
However, we find that the resulting solutions are still close to periodic and one of the two frequencies $\omega_j$ turns out to be small. After setting the smallest $\omega_j$ to zero for each of them, we recompute the spectra with the forward NFT.
They remain similar and symmetric as shown in (b). The resulting waveform is quasi-periodic, but no longer genus-2. Spurious eigenvalues emerge close to the real axis, which can be viewed as a small perturbation of the genus-2 solution.
Even though the evolution of the waveform is no longer described by~\eqref{eq:thetasol} for one frequency set to zero, the main spectrum will still be constant during propagation. Eigenvalues with small imaginary part can easily be filtered at the receiver.

We note that in general, an arbitrary finite-gap solution can be made periodic by adjusting the $\omega_j$ such that their ratios are rational. Ratios with small integers are desirable. Otherwise a large number of oscillations occur within a period, which would require high sampling rates.

\subsection{Period and group velocity matching}

To match the periods of the periodic solutions corresponding to the spectra in Fig.~\ref{fig:Symbols}(b), we use scaling relations ($s$ is a real scaling factor),
\begin{equation}
\lambda_j\to s\lambda_j,\, \omega_0,\underline{\omega}\to s\omega_0,s\underline{\omega},\, K_0\to s K_0,\, k_0,\underline{k}\to s^2k_0,s^2\underline{k},
\label{eq:scaling}
\end{equation}
which follow from the NLSE~\eqref{eq:nlse} and Eq.~\eqref{eq:periods}.
Making the periods equal results in the spectra shown in Fig.~\ref{fig:Symbols}(c). This costs one real degree of freedom as the overall scale of the spectrum is now determined by the fixed period of the solutions $T_p$.

The group velocities of the resulting symbols differ in general.
One can show that by shifting the frequencies of a solution $q(t)$ to the NLSE by $\Delta\omega$ through a phase factor, $q(t) \to q(t)\exp(i \Delta\omega\, t)$, the function $q(t-2\Delta\omega z,z)\exp[i\Delta\omega(t-\Delta\omega z)]$ also fulfills the NLSE, however in a reference frame that moves in time relative to the original solution (see, for example,~\cite[Sec. 3.10.2]{Kremp2011}).
A frequency shift therefore changes the group velocity of a symbol. From~\eqref{eq:quasi} it follows that a shift in the linear spectrum corresponds to a shift in the nonlinear spectrum. We use this degree of freedom to match the group velocities. 
The resulting spectra are shown in Fig.~\ref{fig:Symbols}(d). While the initial nonlinear spectra are symmetric, enforcing the constraints leads to an irregular constellation.

The waveforms in the spectrum are labeled according to the legend in Fig.~\ref{fig:Symbols}(d). This labeling ensures that the closest symbols according to the metric of Eq.~\eqref{eq:metric} below differ by only one bit.

\subsection{Signal synthesis}

Provided that the envelope of the waveform evolves slowly in space over the distance of one span~\cite{Kamalian2017} the link with periodic lumped amplification is described by the dimensionful NLSE with an effective nonlinearity parameter $\gamma_\text{eff} = \gamma[1-\exp(-\alpha L)]/\alpha$. Here $\alpha$ is the attenuation and $L$ the span length.
The physical modulation $A(T,Z)$ of the carrier is obtained from the dimensionless waveform $q(t,z)$ through the relations $Z = (T_0^2/|\beta_2|)z$, $T=T_0t$ and $A(T,Z) = \sqrt{|\beta_2|/(\gamma_\text{eff} T_0)}q(t,z)$. Here $\beta_2$ is the group velocity dispersion constant and $T_0$ can be chosen freely to set the time period of the symbols. We choose $T_0$ to obtain a symbol duration (2 periods) of $T_p+T_{\text{CP}}=1$ ns which includes a 50\% CP (we state the cyclic prefix length $T_{\text{CP}}$ relative to the total symbol duration). This fixes the power of the signal to $P=2.5$ dBm.
Symbol length, bandwidth $B$, power $P$ and spectral efficiency (SE) of this modulation scheme are summarized in Table~\ref{tab:parameters}.

\begin{table}[h]
\caption{Parameters of the PNFT modulation scheme.}\label{tab:parameters}
\centering
\begin{tabular}{l|l}
Parameter & Value\\
\hline
$T_p + T_{\text{CP}}$ & 1 ns\\
$P $    & 2.5 dBm\\
$B $    & 4.5 GHz\\
SE    & 0.45 bits/s/Hz\\
\end{tabular}
\end{table}

Because the waveforms do not decay to zero, discontinuities between consecutive symbols cannot be avoided. However since the starting point of a symbol can be chosen freely, we set it to the point of minimum amplitude. This minimizes discontinuities at symbol boundaries (as long as an integral number of periods is transmitted) and improves performance as shown below.
Amplitude and phase of the symbols are shown in Fig.~\ref{fig:waveforms} (a) and (b).
Finally we make use of the fact that the spectra for $q(t)$ and $e^{i\phi} q(t)$ are the same to eliminate discontinuities in the phase. Note that this can be done only since the phase is not modulated. Fig.~\ref{fig:waveforms} (c) and (d) show an example of the phase and amplitude of four juxtaposed signals.

\section{Experimental setup}
\label{sec:experiment}

\begin{figure}[b]
\includegraphics[width=0.49\textwidth]{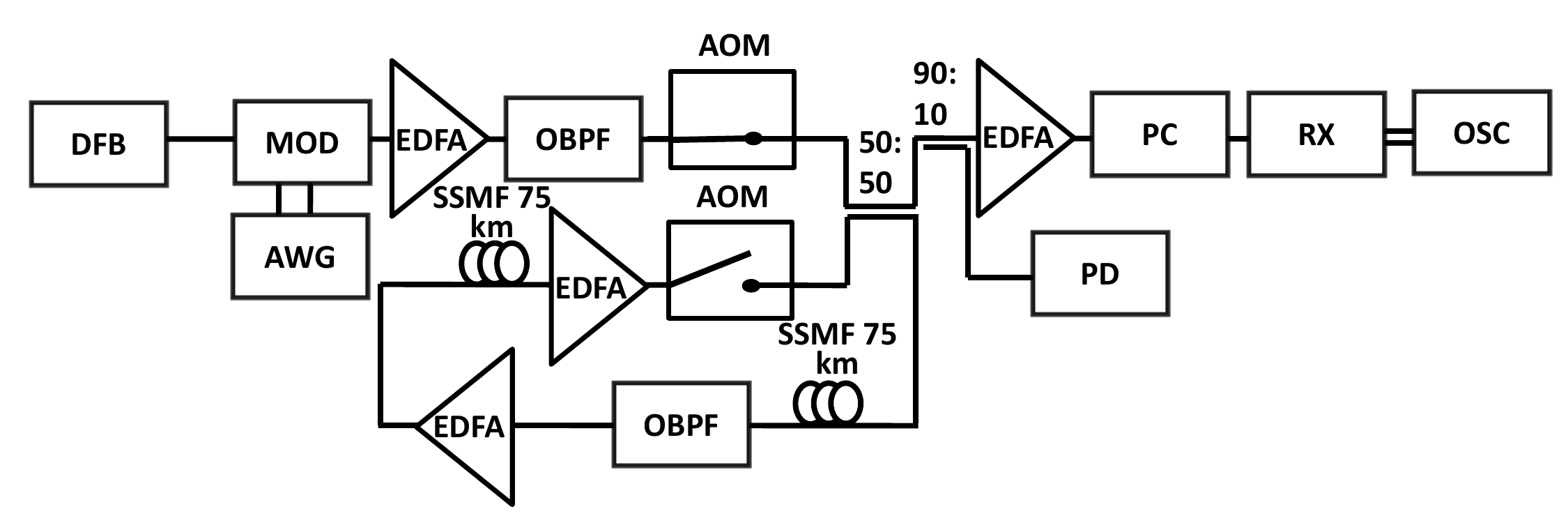}
 \caption{\label{fig:setup} Experimental setup. DFB: Distributed feedback fiber laser; MOD: IQ-modulator, AWG: arbitrary waveform generator; OBPF: optical bandpass filter; EDFA: Erbium-doped fiber amplifier, AOM: acousto-optical modulator; PC: polarization controller; SSMF: standard single mode fiber; PD: Low-bandwidth photodetector; OSC: real-time sampling oscilloscope; RX: balanced receiver.}
\end{figure}

\begin{figure*}[t]
\begin{center}
 \includegraphics[width=0.4\textwidth]{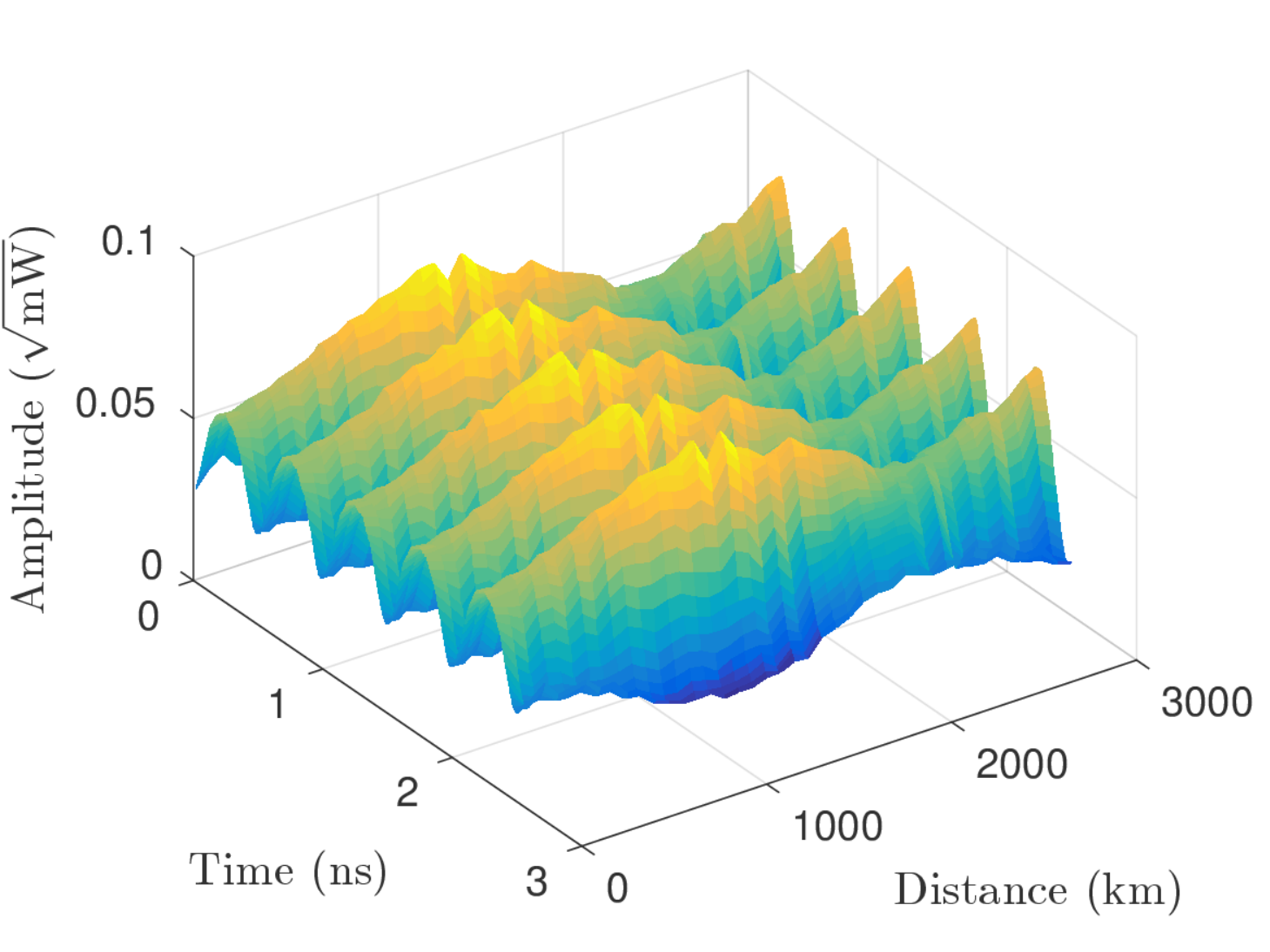} \includegraphics[width=0.4\textwidth]{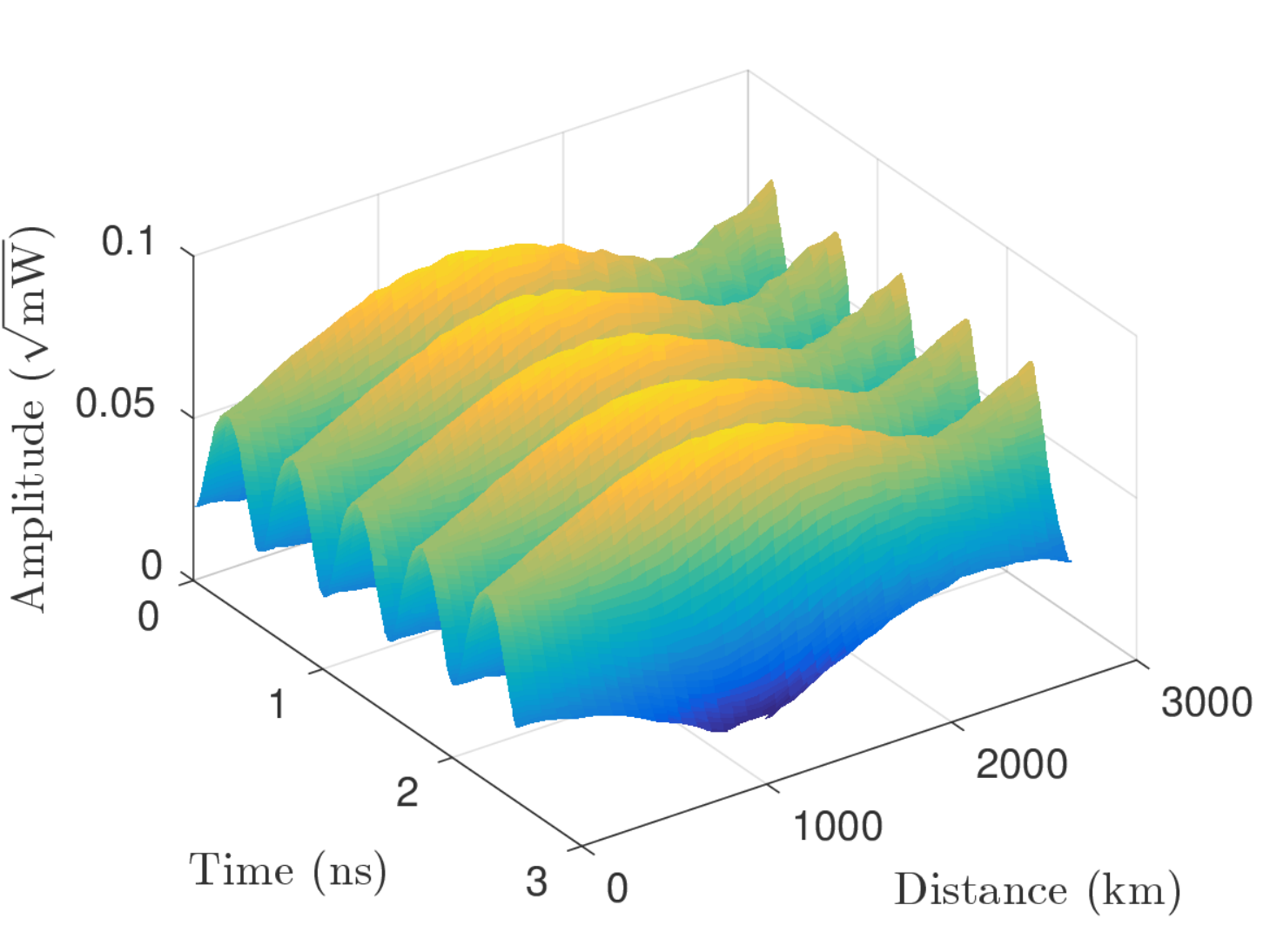}
\end{center}
 \caption{\label{fig:spacetime} Surface plots showing spacetime evolution of the symbol with label '00' (red) of Fig~\ref{fig:Symbols}. Left: Experimentally measured evolution with 75 km spatial resolution. Right: Simulated evolution based on the NLSE without noise and attenuation.}
\end{figure*}

For the experiments, we have employed an optical long-haul transmission testbed including a recirculating fiber loop. We transmit a repeated sequence of PNFT symbols, whose length is limited by the memory of arbitrary waveform generator (AWG).
A train of 2500 PNFT-symbols (2.5 $\mu$s total time duration) is prefixed by single a Schmidl-Cox time synchronization symbol (42 ns) and guard interval (21 ns).

The experimental setup is shown in Fig.~\ref{fig:setup}. The signal is loaded into an AWG operating at 64 Gigasamples/s (GS/s) which corresponds to 32 samples per period. Its output is transferred onto the optical carrier emitted from a single-frequency distributed feedback (DFB) fiber laser (linewidth$\sim 100$ Hz) by a standard IQ-modulator.
To ensure the waveform is not altered by the low-frequency limit of the electrical drivers, we apply a 5 GHz shift to the signal in the digital domain, removing any DC-component.
The signal is amplified, optically filtered and injected into the recirculating fiber loop via a 50:50 coupler. 
The loop operation is controlled by two acousto-optical modulators (AOMs) used as optical switches and driven by an electrical pulse generator. In the shown configuration, multiple sequences of PNFT symbols are injected into the fiber loop until it is filled, which is when the configuration of the AOMs is reversed. A low bandwidth photodetector (PD) is used to ensure the power of the signal inside the loop is stable. After each fiber loop, the signal is coupled out of the loop and amplified. We gather the power into a single polarization through the polarization controller before it is detected by an intradyne balanced receiver and acquired with a real-time oscilloscope at 80 GS/s. The oscilloscope is synchronized with the AOMs through the pulse generator and allows us to record data after a specified number of revolutions. 
Digital signal processing is performed in software and consists of timing synchronization, downsampling to 64 GS/s, compensation of the deterministic carrier frequency offset (CFO) caused by the AOMs, and the forward PNFT. We do not compensate stochastic CFO, because it is negligible on the time-scale of the symbols.

\section{Results}
\label{sec:results}

\begin{figure*}[t]
\centering
	\includegraphics[width=0.4\textwidth]{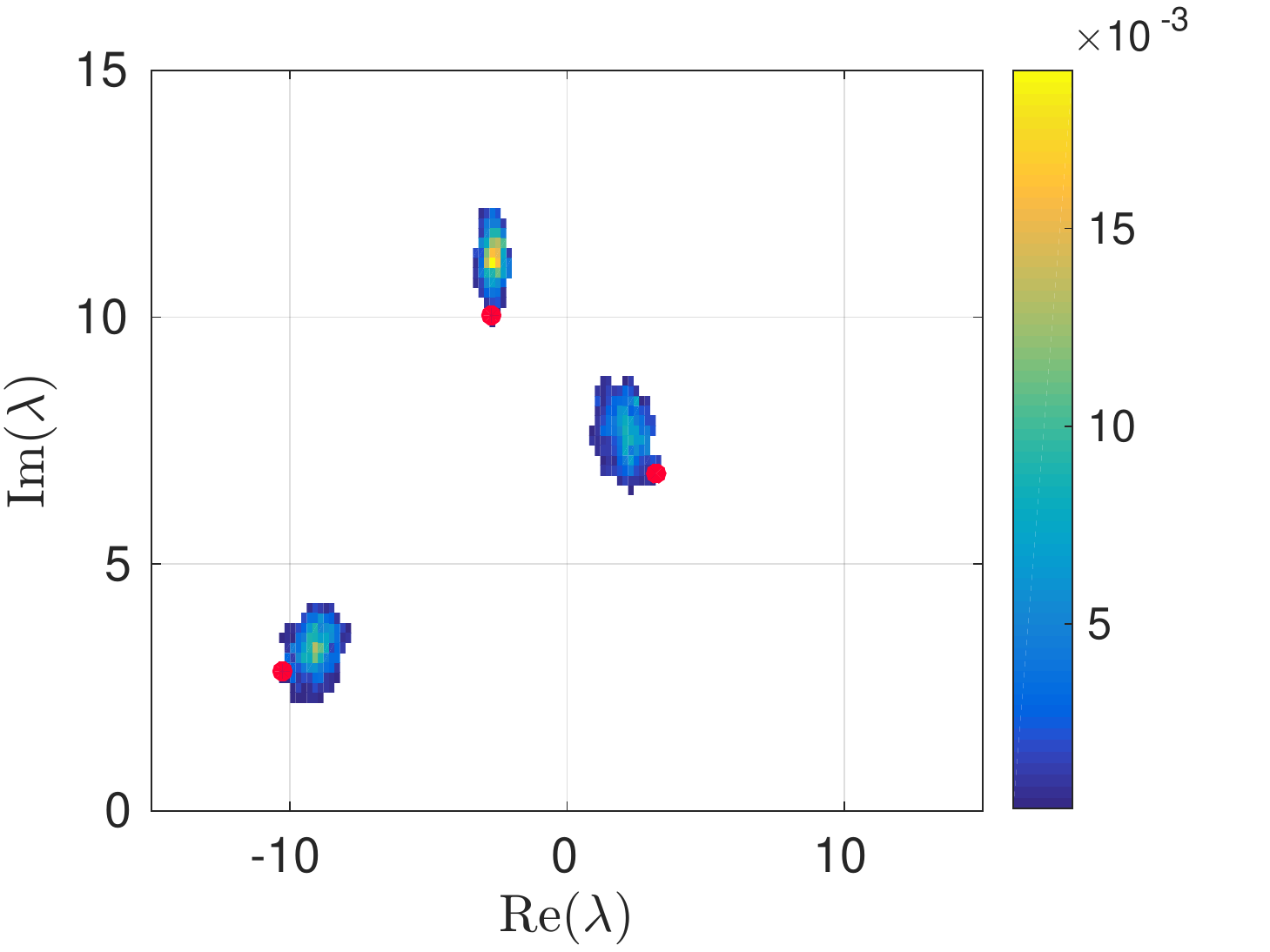} \includegraphics[width=0.4\textwidth]{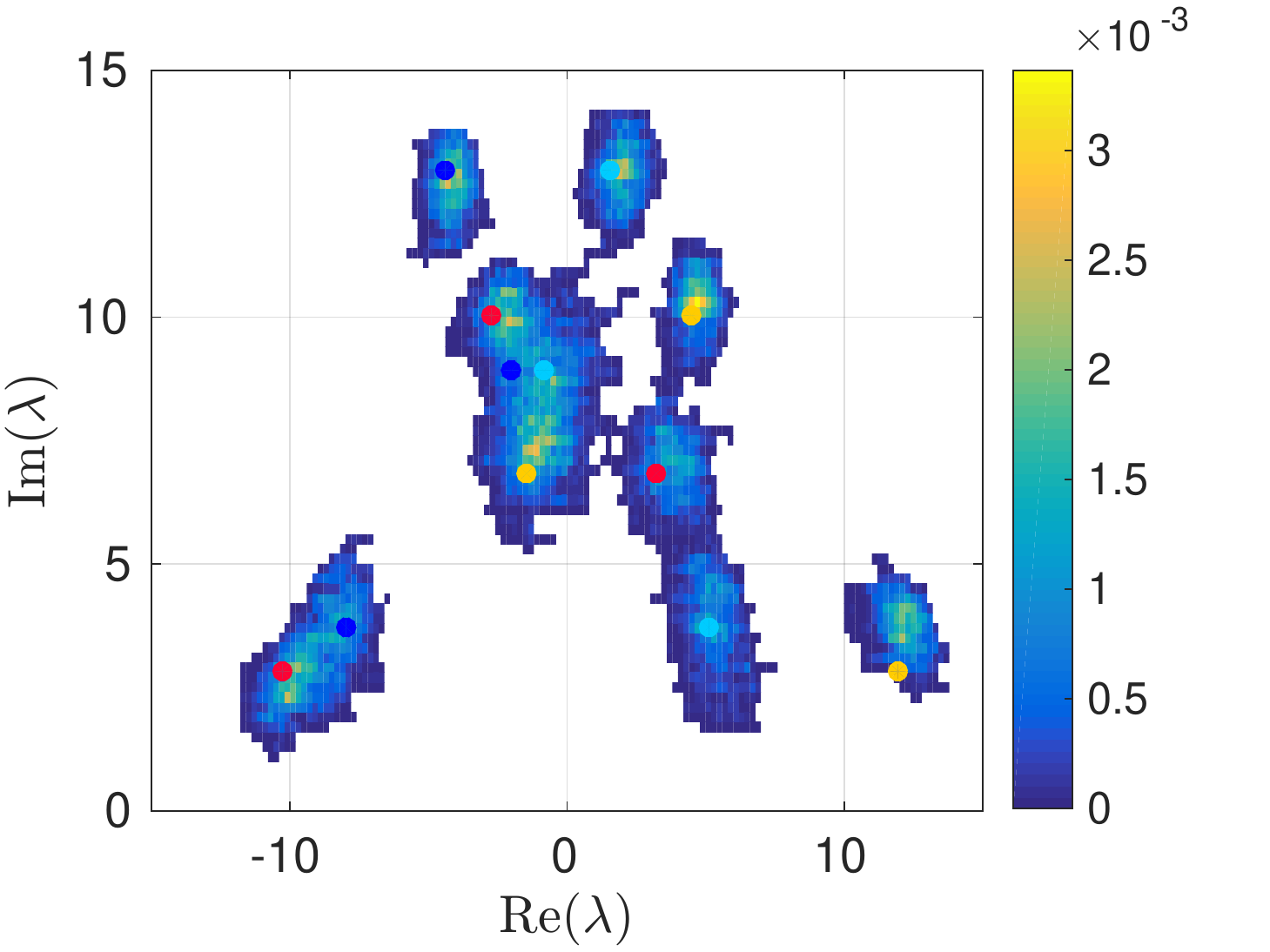}
\caption{\label{fig:noise_distribution}  Left: Single signal transmission after 3000 km (10000 symbols). Right: Received constellation after 2250 km (10000 symbols). Colored circles mark the position of the original signal points.}
\end{figure*}

We first study the spacetime evolution of one of the periodic waveforms. To this end we operate the experimental setup with a single fiber span and EDFA in the recirculating loop. To stabilize the power fluctuations due to polarization dependent losses, an extra polarization controller is included inside the loop. By performing measurements for a varying number of revolutions we map the nonlinear spacetime evolution of the waveform with a spatial resolution of 75 km.
We transmitted 2500 periods of the same waveform corresponding to the symbol with label '00' (red) shown in Fig.~\ref{fig:waveforms}, with the corresponding main spectrum shown in Fig.~\ref{fig:Symbols}
For data analysis we discard 100 periods on each end of the pulse train at the receiver to avoid boundary effects.
The remaining data are grouped into 460 bunches of 5 periods, and averaged to reduce the influence of amplifier noise. 

Fig.~\ref{fig:spacetime} compares the experimentally measured spacetime evolution to split-step Fourier method simulation results for the noise-less effective model over a distance of 3000 km (40 spans). We find good qualitative agreement. The spatial dynamics are slow compared to the span length as required for the applicability of the effective model. SSFM simulations suggest the resulting waveform is approximately (quasi-)periodic in space.

In the left panel of Fig.~\ref{fig:noise_distribution} we show a heatmap of the associated nonlinear spectrum. It is obtained by performing the forward PNFT for each period of the received waveform separately after upsampling to 1024 samples per symbol and collecting the result in a histogram. For the forward transform we employ the fast forward PNFT algorithm provided in the FNFT-package~\cite{Wahls2018}.
We observe three well-defined clusters close to the original spectral points. The distributions are mainly due to the amplifier noise and the different eigenvalues in a spectrum are correlated.
In simulations without noise, we find that the combination of fiber attenuation and periodic amplification   lead to small periodic oscillations of the spectrum  as a function of distance.  Further simulations suggest that the observed systematic  displacement is explained by the Gaussian-shaped optical filter in the loop. In case it is not ideally centered, the (linear) spectrum is slightly distorted, leading to the observed deviations~\cite{Goossens2019a}.

\begin{figure}[ht]
 \center{\includegraphics[width=0.5\textwidth]{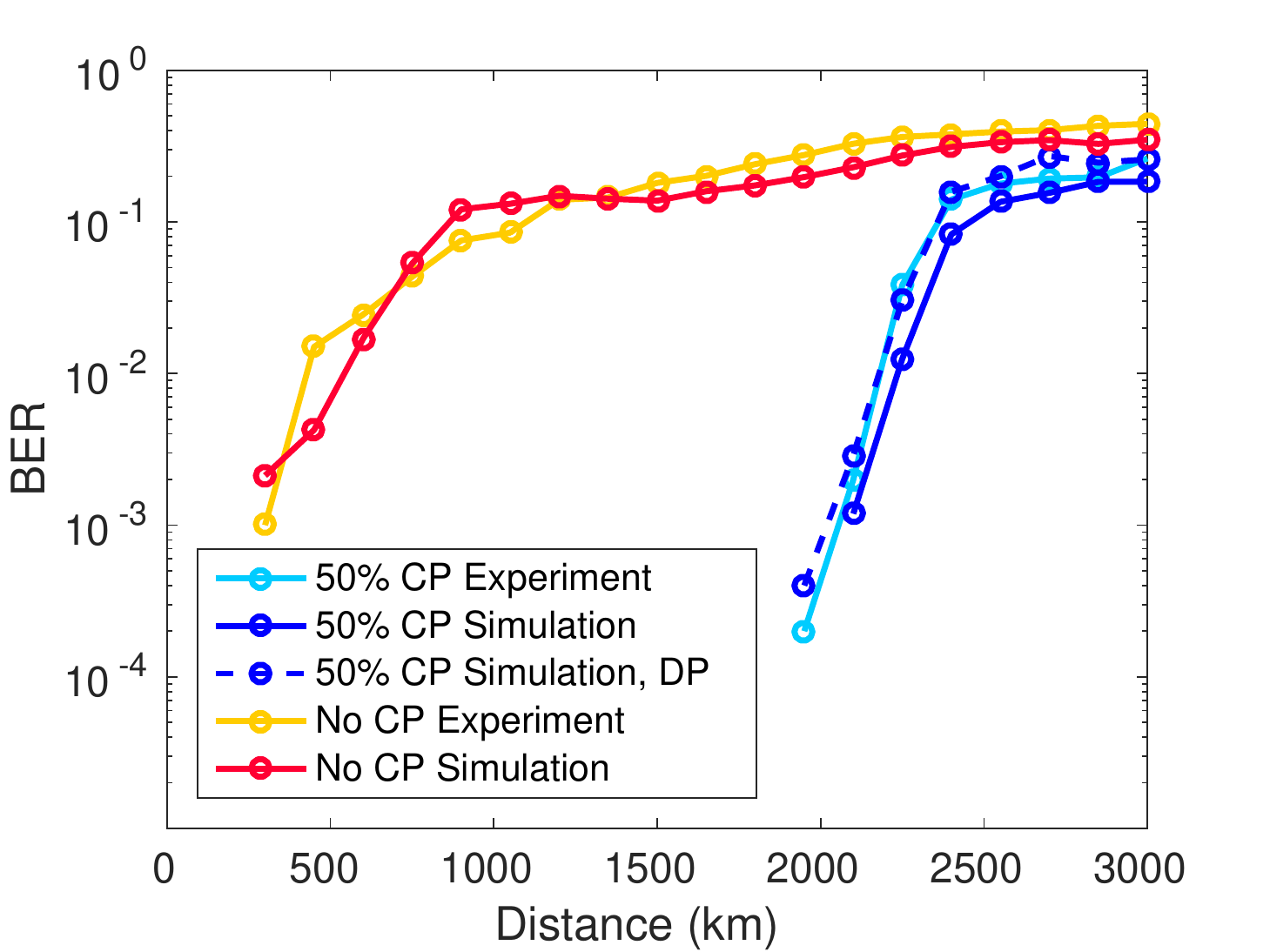}}
 \caption{\label{fig:experiment_ber} BER-curves for 50\% CP and without CP in experiment and simulations of the NLSE including noise and attenuation. DP stands for discontinuous phase. Enforcing a continuous phase at symbol boundaries provides a gain of approximately 70 km, or approximately 1 span.}
\end{figure}

Next we perform a data transmission experiment. To this end we reconfigure the loop to accommodate two fiber spans.
For the analysis 100 periods are again discarded at each end of the pulse train of 2500 periods to avoid boundary effects. The apparent loss in spectral efficiency is an artifact because the  transmitted pulse train is finite and would not occur in a real transmission system.
Demapping to obtain the received constellation symbol $\hat{s}$ for the computation of the bit error ratio (BER) is based on the following metric:
\begin{equation}
\hat{s} = \arg \min_{s=1\ldots 4} \left(\min_{\text{perm}(\lambda_k^{(s)})} \sum_{k=1}^{g+1}|\lambda_{k}^{(s)} - \lambda^{\text{recv}}_k|^2\right).
\label{eq:metric}
\end{equation}
The $\lambda_k^{(s)}$ denote the four spectra in the constellation, while $\lambda_k^{\text{recv}}$ is the received spectrum that is being demapped.
The metric takes into account that the three spectrum points of each symbol are indistinguishable.
The BER-curve, measured over 4 signal-bursts totaling $10^4$ symbols is shown in Fig.~\ref{fig:experiment_ber}. We obtain a BER $< 10^{-3}$ up to a distance of 2000 km. We further observe good agreement between simulation of the NLSE including amplifier noise and the experiment. 

Fig.~\ref{fig:noise_distribution} shows the obtained constellation points after 2250 km. Note that here the forward PNFT is computed after removing the CP for each symbol. The spurious eigenvalues are removed from the main spectrum by keeping only the three points with largest imaginary part. 
The broadening of the distributions of main spectral points is considerably enhanced compared to the transmission of a single symbol, which suggests ISI as the cause.

To analyze the impact of ISI, note that in the absence of nonlinearity, the dispersive temporal broadening of a (Gaussian) pulse can be estimated from its bandwidth $B$, propagation distance $z$, and the dispersion parameter $\beta_2$ as:
\begin{equation}
\Delta T=2\pi|\beta_2|z B.\label{eq:memory}
\end{equation}
CPs of neighboring symbols start overlapping when the broadening $\Delta T$ exceeds the CP length: $\Delta T> T_\text{CP}$.
With a 99\%~bandwidth of $B=4.5$ GHz at the transmitter, $\beta_2= -21.5\ \textrm{ps}^2\textrm{km}^{-1}$ and $\Delta T = T_\text{CP}=0.5$ ns we would expect interference-free propagation up to 900 km. 

Fig.~\ref{fig:experiment_ber} shows that without CP, a BER of $10^{-3}$ is exceeded already after about 300 km. Therefore one would roughly expect the same threshold to be exceeded after approximately 1200 km when including CP. However we observe a much larger distance of 2000 km for this threshold.
A possible explanation is that Eq.~\eqref{eq:memory} is not applicable because nonlinearity influences the evolution of the waveform in an essential way (it is clearly violated for solitons). Indeed one can see from Fig.~\ref{fig:spacetime} that the waveform initially exhibits temporal focusing, reaching maximum temporal compression (maximal spectral broadening) after $Z=1275$ km. After this point temporal broadening sets in again. 
We can compute the zero-boundary NFT of a single period  placed on a zero background to gain insight into the solitonic character of the waveform. The resulting discrete spectrum consists of a single point which contains $\sim 90$\% of the energy of the waveform, which can hence be characterized as soliton-like.

Finally in the same figure we also show a simulation of 50\% CP without making the phase between consecutive symbols continuous at the transmitter. We observe that the reach for this system is approximately 70 km less.

\begin{figure}
 \center{\includegraphics[width=0.5\textwidth]{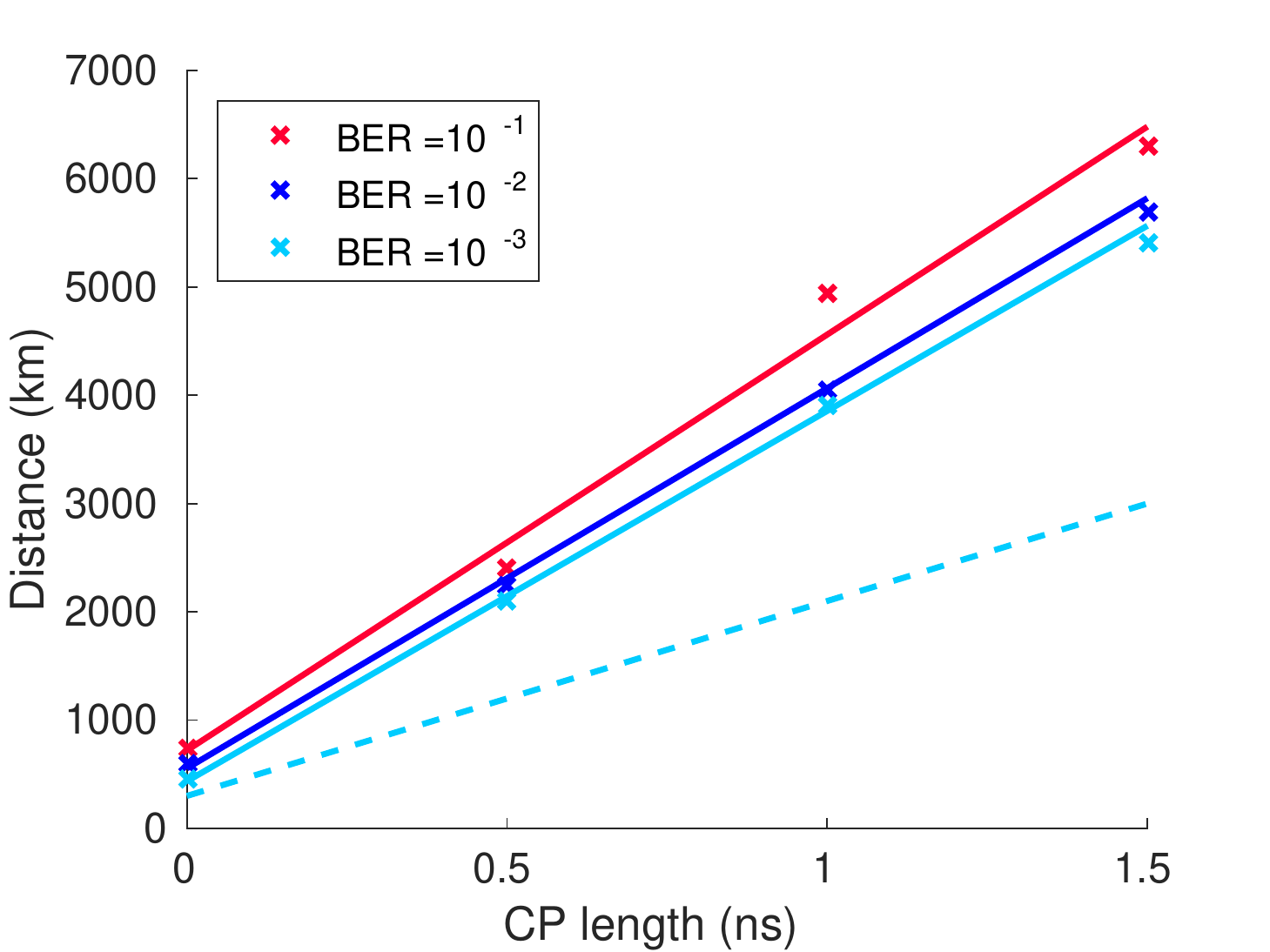}}
 \caption{\label{fig:simulation_reach} Simulation results showing transmission reach as a function of CP length (in integer multiples of the period) for different BER-thresholds. The lines are linear fits to the data and suggest a linear relationship between CP length and reach. The dashed line is an estimate based on dispersive broadening alone.}
\end{figure}

To determine the effect of the CP length on the transmission reach, we simulate data-transmission of $10^5$ symbols over a lossy fiber for four different CP lengths: $0$ ns, $0.5$ ns (50\%), $1$ ns (66\%) and 1.5 ns (75\%). At each amplifier white Gaussian noise is added assuming a noise figure of $5.5$ dB. We determine the reach for BER-thresholds of $10^{-3}$, $10^{-2}$ and $10^{-1}$. The symbols for different CP lengths are designed by minimizing the discontinuity at the boundary, as before. 
The result is plotted in Fig.~\ref{fig:simulation_reach}.  We observe that the reach grows approximately linearly with CP, but much faster than predicted based on group velocity dispersion, Eq.~\eqref{eq:memory}, showing the solitonic component of the waveforms increases the reach significantly.

\section{Discussion}
\label{sec:discussion}
The method we have proposed in this paper for designing constellations of periodic waveforms with the same finite period and group velocity is based on the exact inverse periodic nonlinear Fourier transform developed in Part I~\cite{GoossensI}.
In a proof of concept, successful transmission with a BER below $10^{-3}$ over a distance of 2000 km at a data rate of 2 Gbs has been demonstrated both experimentally and in simulation.
The large solitonic component of the resulting periodic waveforms allows transmission over distances longer than naively expected based on the cyclic prefix length and bandwidth of the signal.
The waveforms can be regarded as \emph{dressed solitons}. They generalize the concept of eigenvalue communication to periodic waveforms.

The transmission distance of 2000 km is unprecedented for PNFT-based transmission. The achieved spectral efficiency however is still quite low (0.45 bits/s/Hz). Spectral efficiencies of 2.5 bits/s/Hz~\cite{Kopae2020} and 2.8 bits/s/Hz~\cite{Kamalian2017b} have been reported.
Nonlinear frequency amplitude modulation proposed in Part I achieves 0.67 bits/s/Hz (over 1500 km), which is close to the 0.75 bits/s/Hz achieved in one-soliton transmission (over 1000 km)~\cite{Gui2017}. NFAM can be extended to higher spectral efficiencies.

To make PNFT transmission attractive, it is necessary to significantly increase spectral efficiency. By increasing the genus and the length of the signal (period), the number of bits per symbol can be increased while keeping the bandwidth and hence the cyclic prefix length constant.
In addition, all available degrees of freedom should be modulated.
Here we encoded information on two real degrees of freedom of a genus 2 spectrum (six total), while two were used to control the power and group velocity of the signal and two remained unused.
For a waveform of genus $g$, the auxiliary spectrum provides $g$ additional real degrees of freedom. The constellation design is an open problem.
Spectral efficiency can be further doubled by polarization-division multiplexing and has been implemented for transmission based on the conventional NFT. The Manakov system describing the polarized optical field is integrable~\cite{Manakov1974}, but only very preliminary results are available so far for the generalization of the periodic NFT to two polarizations~\cite{Christiansen2000,Adams1993}.

To design systems, a deeper theoretical understanding is required. This includes a channel model that accounts for the correlated non-Gaussian noise. A nonlinear analogue of the Nyquist-Shannon sampling theorem would allow a more efficient (albeit signal-dependent) signal sampling. To fully leverage the potential of NFT-based transmission a method for nonlinear multiplexing is required, which ideally can be performed all-optically.

\bibliographystyle{IEEEtran}
\bibliography{GoossensPartII}

\begin{thebibliography}{10}
\providecommand{\url}[1]{#1}
\csname url@samestyle\endcsname
\providecommand{\newblock}{\relax}
\providecommand{\bibinfo}[2]{#2}
\providecommand{\BIBentrySTDinterwordspacing}{\spaceskip=0pt\relax}
\providecommand{\BIBentryALTinterwordstretchfactor}{4}
\providecommand{\BIBentryALTinterwordspacing}{\spaceskip=\fontdimen2\font plus
\BIBentryALTinterwordstretchfactor\fontdimen3\font minus
  \fontdimen4\font\relax}
\providecommand{\BIBforeignlanguage}[2]{{%
\expandafter\ifx\csname l@#1\endcsname\relax
\typeout{** WARNING: IEEEtran.bst: No hyphenation pattern has been}%
\typeout{** loaded for the language `#1'. Using the pattern for}%
\typeout{** the default language instead.}%
\else
\language=\csname l@#1\endcsname
\fi
#2}}
\providecommand{\BIBdecl}{\relax}
\BIBdecl

\bibitem{Richardson2013}
D.~Richardson, J.~Fini, and L.~Nelson, ``Space-division multiplexing in optical
  fibres,'' \emph{Nature Photonics}, vol.~7, no.~5, pp. 354--362, 2013.

\bibitem{Bocherer2019}
G.~{Böcherer}, P.~{Schulte}, and F.~{Steiner}, ``Probabilistic shaping and
  forward error correction for fiber-optic communication systems,''
  \emph{Journal of Lightwave Technology}, vol.~37, no.~2, pp. 230--244, Jan
  2019.

\bibitem{Ellis2010}
\BIBentryALTinterwordspacing
A.~D. Ellis, J.~Zhao, and D.~Cotter, ``Approaching the non-linear {S}hannon
  limit,'' \emph{J. Lightwave Technol.}, vol.~28, no.~4, pp. 423--433, Feb
  2010. [Online]. Available:
  \url{http://jlt.osa.org/abstract.cfm?URI=jlt-28-4-423}
\BIBentrySTDinterwordspacing

\bibitem{Secondini2016}
M.~{Secondini} and E.~{Forestieri}, ``The limits of the nonlinear {S}hannon
  limit,'' in \emph{2016 Optical Fiber Communications Conference and Exhibition
  (OFC)}, March 2016, pp. 1--3.

\bibitem{Essiambre2010}
R.-J. Essiambre, G.~Kramer, P.~J. Winzer, G.~J. Foschini, and B.~Goebel,
  ``Capacity limits of optical fiber networks,'' \emph{Journal of Lightwave
  Technology}, vol.~28, no.~4, pp. 662--701, 2010.

\bibitem{Turitsyn2017}
S.~K. Turitsyn, J.~E. Prilepsky, S.~T. Le, S.~Wahls, L.~L. Frumin, M.~Kamalian,
  and S.~A. Derevyanko, ``Nonlinear fourier transform for optical data
  processing and transmission: advances and perspectives,'' \emph{Optica},
  vol.~4, no.~3, pp. 307--322, 2017.

\bibitem{Hasegawa1993}
A.~Hasegawa and T.~Nyu, ``Eigenvalue communication,'' \emph{Journal of
  lightwave technology}, vol.~11, no.~3, pp. 395--399, 1993.

\bibitem{Yousefi2014}
M.~I. Yousefi and F.~R. Kschischang, ``Information transmission using the
  nonlinear {F}ourier transform, {P}art {I}-{III},'' \emph{IEEE Transactions on
  Information Theory}, vol.~60, no.~7, pp. 4312--4369, 2014.

\bibitem{Le2014}
\BIBentryALTinterwordspacing
S.~T. Le, J.~E. Prilepsky, and S.~K. Turitsyn, ``Nonlinear inverse synthesis
  for high spectral efficiency transmission in optical fibers,'' \emph{Opt.
  Express}, vol.~22, no.~22, pp. 26\,720--26\,741, Nov 2014. [Online].
  Available: \url{http://www.opticsexpress.org/abstract.cfm?URI=oe-22-22-26720}
\BIBentrySTDinterwordspacing

\bibitem{Civelli2017}
S.~{Civelli}, E.~{Forestieri}, and M.~{Secondini}, ``Why noise and dispersion
  may seriously hamper nonlinear frequency-division multiplexing,'' \emph{IEEE
  Photonics Technology Letters}, vol.~29, no.~16, pp. 1332--1335, Aug 2017.

\bibitem{Wahls2015}
S.~Wahls and H.~V. Poor, ``Fast numerical nonlinear {F}ourier transforms,''
  \emph{IEEE Transactions on Information Theory}, vol.~61, no.~12, pp.
  6957--6974, 2015.

\bibitem{Vaibhav2018}
V.~Vaibhav, ``Higher order convergent fast nonlinear {F}ourier transform,''
  \emph{IEEE Photonics Technology Letters}, vol.~30, no.~8, pp. 700--703, 2018.

\bibitem{Chimmalgi2018}
S.~Chimmalgi, P.~Prins, and S.~Wahls, ``\BIBforeignlanguage{English}{Fast
  nonlinear {F}ourier transform algorithms using higher order exponential
  integrators},'' \emph{\BIBforeignlanguage{English}{IEEE Access}}, vol.~7,
  no.~1, pp. 145\,161--145\,176, 2019.

\bibitem{Kamalian2017}
M.~Kamalian, J.~E. Prilepsky, S.~T. Le, and S.~K. Turitsyn, ``On the design of
  {NFT}-based communication systems with lumped amplification,'' \emph{Journal
  of Lightwave Technology}, vol.~35, no.~24, pp. 5464--5472, 2017.

\bibitem{Goossens2017}
\BIBentryALTinterwordspacing
J.-W. Goossens, M.~I. Yousefi, Y.~Jaou\"{e}n, and H.~Hafermann,
  ``Polarization-division multiplexing based on the nonlinear {F}ourier
  transform,'' \emph{Opt. Express}, vol.~25, no.~22, pp. 26\,437--26\,452, Oct
  2017. [Online]. Available:
  \url{http://www.opticsexpress.org/abstract.cfm?URI=oe-25-22-26437}
\BIBentrySTDinterwordspacing

\bibitem{Gaiarin2018}
\BIBentryALTinterwordspacing
S.~Gaiarin, A.~M. Perego, E.~P. da~Silva, F.~D. Ros, and D.~Zibar,
  ``Dual-polarization nonlinear {F}ourier transform-based optical communication
  system,'' \emph{Optica}, vol.~5, no.~3, pp. 263--270, Mar 2018. [Online].
  Available:
  \url{http://www.osapublishing.org/optica/abstract.cfm?URI=optica-5-3-263}
\BIBentrySTDinterwordspacing

\bibitem{Civelli2018}
\BIBentryALTinterwordspacing
S.~Civelli, S.~K. Turitsyn, M.~Secondini, and J.~E. Prilepsky,
  ``Polarization-multiplexed nonlinear inverse synthesis with standard and
  reduced-complexity nft processing,'' \emph{Opt. Express}, vol.~26, no.~13,
  pp. 17\,360--17\,377, Jun 2018. [Online]. Available:
  \url{http://www.opticsexpress.org/abstract.cfm?URI=oe-26-13-17360}
\BIBentrySTDinterwordspacing

\bibitem{Yousefi2020}
M.~{Yousefi} and X.~{Yangzhang}, ``Linear and nonlinear frequency-division
  multiplexing,'' \emph{IEEE Transactions on Information Theory}, vol.~66,
  no.~1, pp. 478--495, Jan 2020.

\bibitem{Le2017}
S.~T. Le, V.~Aref, and H.~Buelow, ``High speed precompensated nonlinear
  frequency-division multiplexed transmissions,'' \emph{Journal of Lightwave
  Technology}, vol.~36, no.~6, pp. 1296--1303, 2017.

\bibitem{Le2017b}
S.~T. Le and H.~Buelow, ``64$\times$ 0.5 {G}baud nonlinear frequency division
  multiplexed transmissions with high order modulation formats,'' \emph{Journal
  of Lightwave Technology}, vol.~35, no.~17, pp. 3692--3698, 2017.

\bibitem{Le2018}
S.~T. Le, K.~Schuh, F.~Buchali, and H.~Buelow, ``100 {G}bps b-modulated
  nonlinear frequency division multiplexed transmission,'' in \emph{Optical
  Fiber Communication Conference}.\hskip 1em plus 0.5em minus 0.4em\relax
  Optical Society of America, 2018, pp. W1G--6.

\bibitem{Aref2018}
V.~Aref, S.~T. Le, and H.~Buelow, ``Modulation over nonlinear {F}ourier
  spectrum: Continuous and discrete spectrum,'' \emph{Journal of Lightwave
  Technology}, vol.~36, no.~6, pp. 1289--1295, 2018.

\bibitem{Gaiarin2018a}
S.~{Gaiarin}, F.~{Da Ros}, N.~{De Renzis}, E.~P. {da Silva}, and D.~{Zibar},
  ``Dual-polarization {NFDM} transmission using distributed {R}aman
  amplification and {NFT}-domain equalization,'' \emph{IEEE Photonics
  Technology Letters}, vol.~30, no.~22, pp. 1983--1986, Nov 2018.

\bibitem{Gui2018}
T.~Gui, G.~Zhou, C.~Lu, A.~P.~T. Lau, and S.~Wahls, ``Nonlinear frequency
  division multiplexing with b-modulation: shifting the energy barrier,''
  \emph{Optics express}, vol.~26, no.~21, pp. 27\,978--27\,990, 2018.

\bibitem{Gui2018a}
T.~{Gui}, W.~A. {Gemechu}, J.~{Goossens}, M.~{Song}, S.~{Wabnitz}, M.~I.
  {Yousefi}, H.~{Hafermann}, A.~P.~T. {Lau}, and Y.~{Jaouën},
  ``Polarization-division-multiplexed nonlinear frequency division
  multiplexing,'' in \emph{2018 Conference on Lasers and Electro-Optics
  (CLEO)}, May 2018, pp. 1--2.

\bibitem{Yangzhang2019}
X.~Yangzhang, S.~T. Le, V.~Aref, H.~Buelow, D.~Lavery, and P.~Bayvel,
  ``Experimental demonstration of dual-polarisation {NFDM} transmission with
  b-modulation,'' \emph{IEEE Photonics Technology Letters}, 2019.

\bibitem{Belokolos1994}
E.~D. Belokolos, \emph{Algebro-geometric approach to nonlinear integrable
  equations}.\hskip 1em plus 0.5em minus 0.4em\relax Springer, 1994.

\bibitem{Smirnov2013}
A.~O. Smirnov, ``Periodic two-phase ``rogue waves'','' \emph{Mathematical
  Notes}, vol.~94, no. 5-6, pp. 897--907, 2013.

\bibitem{Kamalian2016}
M.~Kamalian, J.~E. Prilepsky, S.~T. Le, and S.~K. Turitsyn, ``Periodic
  nonlinear {F}ourier transform for fiber-optic communications, {P}art {I}:
  theory and numerical methods,'' \emph{Optics express}, vol.~24, no.~16, pp.
  18\,353--18\,369, 2016.

\bibitem{GoossensI}
J.-W. Goossens, H.~Hafermann, and Y.~Jaou{\"e}n, ``Data transmission based on
  exact inverse periodic nonlinear {F}ourier transform, part {I}: {T}heory,''
  \emph{Journal of Lightwave Technology}, 2020.

\bibitem{Kotlyarov2014}
V.~Kotlyarov and A.~Its, ``Periodic problem for the nonlinear {S}chr{\"o}dinger
  equation,'' \emph{{\rm arXiv preprint} arXiv:1401.4445}, 2014.

\bibitem{Tracy1984}
E.~R. Tracy, ``Topics in nonlinear wave theory with applications,'' Ph.D.
  dissertation, University of Maryland, 1984.

\bibitem{Tracy1988}
E.~Tracy and H.~Chen, ``Nonlinear self-modulation: An exactly solvable model,''
  \emph{Phys. Rev. A}, vol.~37, no.~3, pp. 815--839, 1988.

\bibitem{Kamalian2016a}
M.~Kamalian, J.~E. Prilepsky, S.~T. Le, and S.~K. Turitsyn, ``Periodic
  nonlinear {F}ourier transform for fiber-optic communications, {P}art {II}:
  eigenvalue communication,'' \emph{Opt. Express}, vol.~24, no.~16, pp.
  18\,370--18\,381, 2016.

\bibitem{Kamalian2018c}
M.~Kamalian, A.~Vasylchenkova, J.~Prilepsky, D.~Shepelsky, and S.~Turitsyn,
  ``Communication system based on periodic nonlinear {F}ourier transform with
  exact inverse transformation,'' in \emph{2018 European Conference on Optical
  Communication (ECOC)}.\hskip 1em plus 0.5em minus 0.4em\relax IEEE, 2018, pp.
  1--3.

\bibitem{Kamalian2018}
M.~Kamalian, A.~Vasylchenkova, D.~Shepelsky, J.~E. Prilepsky, and S.~K.
  Turitsyn, ``Signal modulation and processing in nonlinear fibre channels by
  employing the {R}iemann--{H}ilbert problem,'' \emph{Journal of Lightwave
  Technology}, vol.~36, no.~24, pp. 5714--5727, 2018.

\bibitem{Kamalian2018b}
M.~Kamalian, J.~Prilepsky, A.~Vasylchenkova, D.~Shepelsky, and S.~Turitsyn,
  ``Methods of nonlinear {F}ourier-based optical transmission with
  periodically-extended signals,'' in \emph{2018 IEEE International Conference
  on the Science of Electrical Engineering in Israel (ICSEE)}.\hskip 1em plus
  0.5em minus 0.4em\relax IEEE, 2018, pp. 1--5.

\bibitem{Kopae2020}
M.~K. Kopae, A.~Vasylchenkova, D.~Shepelsky, J.~E. Prilepsky, and S.~K.
  Turitsyn, ``Full-spectrum periodic nonlinear fourier transform optical
  communication through solving the riemann-hilbert problem,'' \emph{Journal of
  Lightwave Technology}, 2020.

\bibitem{Goossens2019}
J.-W. Goossens, Y.~Jaou{\"e}n, and H.~Hafermann, ``Experimental demonstration
  of data transmission based on the exact inverse periodic nonlinear {F}ourier
  transform,'' in \emph{Optical Fiber Communication Conference}.\hskip 1em plus
  0.5em minus 0.4em\relax Optical Society of America, 2019, pp. M1I--6.

\bibitem{Agrawal2000}
G.~P. Agrawal, ``Nonlinear fiber optics,'' in \emph{Nonlinear Science at the
  Dawn of the 21st Century}.\hskip 1em plus 0.5em minus 0.4em\relax Springer,
  2000, pp. 195--211.

\bibitem{Ania2004}
J.~D. Ania-Casta{\~n}{\'o}n, ``Quasi-lossless transmission using second-order
  {R}aman amplification and fibre {B}ragg gratings,'' \emph{Optics Express},
  vol.~12, no.~19, pp. 4372--4377, 2004.

\bibitem{Le2015}
S.~T. Le, J.~E. Prilepsky, and S.~K. Turitsyn, ``Nonlinear inverse synthesis
  technique for optical links with lumped amplification,'' \emph{Opt. Express},
  vol.~23, no.~7, pp. 8317--8328, 2015.

\bibitem{Shabat1972}
V.~E. Zakharov and A.~B. Shabat, ``Exact theory of two-dimensional
  self-focusing and one-dimensional self-modulation of waves in nonlinear
  media,'' \emph{Soviet Physics-JETP}, vol.~34, pp. 62--69, 1972.

\bibitem{Ma1981}
Y.-C. Ma and M.~J. Ablowitz, ``The periodic cubic {S}chr{\"o}dinger equation,''
  \emph{Studies in Applied Mathematics}, vol.~65, no.~2, pp. 113--158, 1981.

\bibitem{Wahls2017}
S.~Wahls, ``Generation of time-limited signals in the nonlinear {F}ourier
  domain via b-modulation,'' in \emph{2017 European Conference on Optical
  Communication (ECOC)}.\hskip 1em plus 0.5em minus 0.4em\relax IEEE, 2017, pp.
  1--3.

\bibitem{Wahls2018}
S.~Wahls, S.~Chimmalgi, and P.~J. Prins, ``{FNFT}: a software library for
  computing nonlinear {F}ourier transforms,'' \emph{J. Open Source Software},
  vol.~3, no.~23, pp. 1--11, 2018.

\bibitem{Smirnov2014}
A.~Smirnov, E.~Semenova, V.~Zinger, and N.~Zinger, ``On a periodic solution of
  the focusing nonlinear {S}chr{\"o}dinger equation,'' \emph{{\rm arXiv
  preprint} arXiv:1407.7974}, 2014.

\bibitem{Kremp2011}
T.~Kremp, \emph{Quasi-spectral Finite Difference Methods: Convergence Analysis
  and Application to Nonlinear Optical Pulse Propagation}.\hskip 1em plus 0.5em
  minus 0.4em\relax Cuvillier Verlag, G\"ottingen, 2011.

\bibitem{Goossens2019a}
J.-W. Goossens, H.~Hafermann, and Y.~Jaou{\"e}n, ``Experimental realization of
  {F}ermi-{P}asta-{U}lam-{T}singou recurrence in a long-haul optical fiber
  transmission system,'' \emph{{\rm To appear in} Scientific Reports}, 2019.

\bibitem{Kamalian2017b}
M.~Kamalian, J.~E. Prilepsky, S.~T. Le, and S.~K. Turitsyn, ``Spectral
  efficiency estimation in periodic nonlinear fourier transform based
  communication systems,'' in \emph{2017 Optical Fiber Communications
  Conference and Exhibition (OFC)}.\hskip 1em plus 0.5em minus 0.4em\relax
  IEEE, 2017, pp. 1--3.

\bibitem{Gui2017}
T.~Gui, C.~Lu, A.~P.~T. Lau, and P.~Wai, ``High-order modulation on a single
  discrete eigenvalue for optical communications based on nonlinear fourier
  transform,'' \emph{Optics express}, vol.~25, no.~17, pp. 20\,286--20\,297,
  2017.

\bibitem{Manakov1974}
S.~V. Manakov, ``On the theory of two-dimensional stationary self-focusing of
  electromagnetic waves,'' \emph{Soviet Physics-JETP}, vol.~38, no.~2, pp.
  248--253, 1974.

\bibitem{Christiansen2000}
P.~L. Christiansen, J.~C. Eilbeck, V.~Enolskii, and N.~Kostov,
  ``Quasi--periodic and periodic solutions for coupled nonlinear
  {S}chr{\"o}dinger equations of {M}anakov type,'' \emph{Proceedings of the
  Royal Society of London. Series A: Mathematical, Physical and Engineering
  Sciences}, vol. 456, no. 2001, pp. 2263--2281, 2000.

\bibitem{Adams1993}
M.~Adams, J.~Harnad, and J.~Hurtubise, ``{D}arboux coordinates and
  {L}iouville-{A}rnold integration in loop algebras,'' \emph{Communications in
  mathematical physics}, vol. 155, no.~2, pp. 385--413, 1993.

\end{thebibliography}

\end{document}